\documentclass[a4paper,11pt]{article}
\pdfoutput=1 % if your are submitting a pdflatex (i.e. if you have
\usepackage{jheppub} % for details on the use of the package, please
\usepackage[T1]{fontenc} % if needed

\newcommand\trt{\frac14\textrm{tr}}

\title{\boldmath Higher order fluctuations and correlations of conserved charges from lattice QCD}

\author[a]{Szabolcs Borsanyi,}
\author[a,b,c]{Zoltan Fodor,}
\author[a,d]{Jana N. Guenther} 
\author[d]{Sandor. K. Katz,}
\author[a,c]{K. K. Szab\'o,}
\author[a]{Attila Pasztor,}
\author[e]{Israel Portillo,}
\author[e]{Claudia Ratti}

\affiliation[a]{University  of  Wuppertal, Department  of Physics, Wuppertal  D-42097,  Germany}
\affiliation[b]{E\"otv\"os University, Budapest 1117, Hungary}
\affiliation[c]{J\"ulich  Supercomputing  Centre, J\"ulich  D-52425,  Germany}
\affiliation[d]{University  of  Regensburg, Department of Physics, Regensburg  D-93053,  Germany}
\affiliation[e]{Department  of  Physics,  University  of  Houston, Houston,  TX  77204,  USA}

\emailAdd{borsanyi@uni-wuppertal.de}
\emailAdd{fodor@bodri.elte.hu}
\emailAdd{Jana.Guenther@t-online.de}
\emailAdd{katz@bodri.elte.hu}
\emailAdd{szaboka@general.elte.hu}
\emailAdd{apasztor@bodri.elte.hu}
\emailAdd{iportillovazquez@gmail.com}
\emailAdd{cratti@uh.edu}

\abstract{We calculate several diagonal and non-diagonal fluctuations of
conserved charges in a system of 2+1+1 quark flavors with physical masses, on a
lattice with size $48^3\times12$. Higher order fluctuations at $\mu_B=0$ are
obtained as derivatives of the lower order ones, simulated at imaginary
chemical potential. From these correlations and fluctuations we construct
ratios of net-baryon number cumulants as functions of temperature and chemical
potential, which satisfy the experimental conditions of strangeness neutrality
and proton/baryon ratio. Our results qualitatively explain the behavior of the
measured cumulant ratios by the STAR collaboration.  }

\begin{document} 
\maketitle
\flushbottom

\section{Introduction}
One of the most challenging goals in the study of Quantum Chromodynamics (QCD) is a precise mapping of the phase diagram of strongly interacting matter. First principle, lattice QCD simulations predict that the transition from hadrons to deconfined quarks and gluons is a smooth crossover \cite{Aoki:2006we,Aoki:2006br,Aoki:2009sc,Borsanyi:2010bp,Bhattacharya:2014ara,Bazavov:2011nk}, taking place in the temperature range $T\simeq145-165$~MeV. Lattice simulations cannot presently be performed at finite density due to the sign problem, thus leading to the fact that the QCD phase diagram is still vastly unexplored when the asymmetry between matter and antimatter becomes large.

With the advent of the second Beam Energy Scan (BES-II) at the Relativistic
Heavy Ion Collider (RHIC), scheduled for 2019-2020, there is a renewed interest
in the heavy ion community towards the phases of QCD at moderate-to-large
densities. A rich theoretical effort is being developed in support of the
experimental program; several observables are being calculated, in order to
constrain the existence and location of the QCD critical point and to observe
it experimentally.

Fluctuations of conserved charges (electric charge $Q$, baryon number $B$ and
strangeness $S$) are among the most relevant observables for the finite-density
program for several reasons. One possible way to extend lattice results to
finite density is to perform Taylor expansions of the thermodynamic observables
around chemical potential $\mu_B=0$
\cite{Allton:2002zi,Allton:2005gk,Gavai:2008zr,Basak:2009uv,Kaczmarek:2011zz}.
Fluctuations of conserved charges are directly related to the Taylor expansion
coefficients of such observables, thus, they are needed to extend first
principle approaches to the regions of the phase diagram relevant to RHIC.
An other popular method to extend observables to finite density is the
analytical continuation from imaginary chemical potentials
\cite{Fodor:2001au,deForcrand:2002hgr, DElia:2002tig, Fodor:2001pe, Fodor:2004nz}. The agreement between the analytical continuation and Taylor expansion
was shown for the transition temperature by Bonati et al in Ref.~\cite{Bonati:2018nut}.

Fluctuations can also be measured directly, and a comparison between
theoretical and experimental results allows to extract the chemical freeze-out
temperature $T_f$ and chemical potential $\mu_{Bf}$ as functions of the
collision energy
\cite{Karsch:2012wm,Bazavov:2012vg,Borsanyi:2013hza,Borsanyi:2014ewa,Ratti:2018ksb}. Such fluctuations have been recently calculated and extrapolated
using the Taylor method in Ref.~\cite{Bazavov:2017tot}.
Finally, higher order fluctuations of conserved charges are proportional to
powers of the correlation length and are expected to diverge at the critical
point, thus providing an important signature for its experimental detection
\cite{Gavai:2008zr,Stephanov:1999zu,Cheng:2007jq}.

In this paper, we calculate several diagonal and non-diagonal fluctuations of
conserved charges up to sixth-order and give estimates for higher orders,
in the temperature range $135$~MeV $\leq T\leq$ 220~MeV,
for a system of 2+1+1 dynamical quarks with physical masses and
lattice size $48^3\times12$. We simulate the lower-order fluctuations at
imaginary chemical potential and extract the higher order fluctuations as
derivatives of the lower order ones at $\mu_B=0$. This method has been
successfully used in the past and proved to lead to a more precise
determination of the higher order fluctuations, compared to their direct
calculation \cite{Gunther:2016vcp,DElia:2016jqh}. The direct method (see e.g. \cite{Allton:2002zi})
requires the evaluation of several terms and is affected by a signal-to-noise
ratio which is decreasing as a power law of the spatial volume $V$,  with an
exponent that grows with the order of the susceptibility. 

We also construct combinations of these diagonal and non-diagonal fluctuations
in order to study the ratio of the cumulants of the net-baryon number
distribution as functions of temperature and chemical potential by means of
their Taylor expansion in powers of $\mu_B/T$. We discuss their qualitative
comparison with the experimental results from the STAR collaboration, as well
as the validity of the truncation of the Taylor series.

The paper is organized as follows: we first discuss the use of imaginary
chemical potentials in Section \ref{sec:immu}. Section \ref{sec:ana}
gives details on the lattice setup, on the fitting procedure, on its
generalization for cross-correlators, and finally on the error estimation.
The phenomenological results for the ratios of kurtosis, skewness and variance
of the baryon number are presented in Section \ref{sec:cumulants}.
Conclusions and outlook are discussed in Section \ref{sec:con}, 
while in the Appendix we present all
diagonal and non-diagonal fluctuations needed to construct the cumulant ratios
shown in Section \ref{sec:cumulants}, and give additional technical details.

\section{Fluctuations and imaginary chemical potentials\label{sec:immu}}

The chemical potentials are implemented on a flavor-by-flavor basis, their relation
to the phenomenological baryon ($B$), electric charge ($Q$) and strangeness ($S$) chemical
potentials are given by
\begin{eqnarray}
\mu_u&=&\frac13\mu_B+\frac23\mu_Q
\nonumber\\
\mu_d&=&\frac13\mu_B-\frac13\mu_Q
\nonumber\\
\mu_s&=&\frac13\mu_B-\frac13\mu_Q-\mu_S.
\label{eq:muBQS}
\end{eqnarray}
The observables we are looking at are the derivatives of the free energy with
respect to the chemical potentials. Since the free energy is proportional
to the pressure, we can write:
\begin{equation}
     \chi^{B,Q,S}_{i,j,k}= \frac{\partial^{i+j+k} (p/T^4)}{
(\partial \hat\mu_B)^i
(\partial \hat\mu_Q)^j
(\partial \hat\mu_S)^k
}\,,
\label{eq:chideriv}
\end{equation}
with
\begin{equation}
 \hat\mu_i=\frac{\mu_i}{T}.
\end{equation}
These are the generalized fluctuations we calculated around $\mu=0$ in our previous
work \cite{Bellwied:2015lba}.

The fermion determinant $\det M(\mu)$ is complex for real chemical potentials,
prohibiting the use of traditional simulation algorithms. For imaginary $\mu$,
however, the determinant stays real. The chemical potential is introduced through
weighted temporal links in the staggered formalism:
\begin{eqnarray}
U_0(\mu)=e^\mu U_0,~~~~~~
U_0^\dagger(\mu)=e^{-\mu} U_0^\dagger
\end{eqnarray}
Thus, an imaginary $\mu$ translates into a phase factor for the
antiperiodic boundary condition in the Dirac operator. Due to the $Z(3)$
symmetry of the gauge sector, there is a non-trivial periodicity in
the imaginary quark chemical potential $\mu_q \to \mu_q+i(2\pi/3)T$, which translates
to the baryochemical potential as $\mu_B \to \mu_B+i 2\pi T$, the
Roberge-Weiss symmetry. This is independent of the charge conjugation symmetry 
$\mu_B \leftrightarrow -\mu_B$. As a result, e.g. for the imaginary part of
the baryon density:
\begin{equation}
\left.\langle B\rangle\right|_{\mu_B/T=i\pi-\epsilon}=
-\left.\langle B\rangle\right|_{\mu_B/T=i\pi+\epsilon}
\label{eq:RWconj}
\end{equation}

\begin{figure}[ht]
\begin{center}
\includegraphics[width=3in]{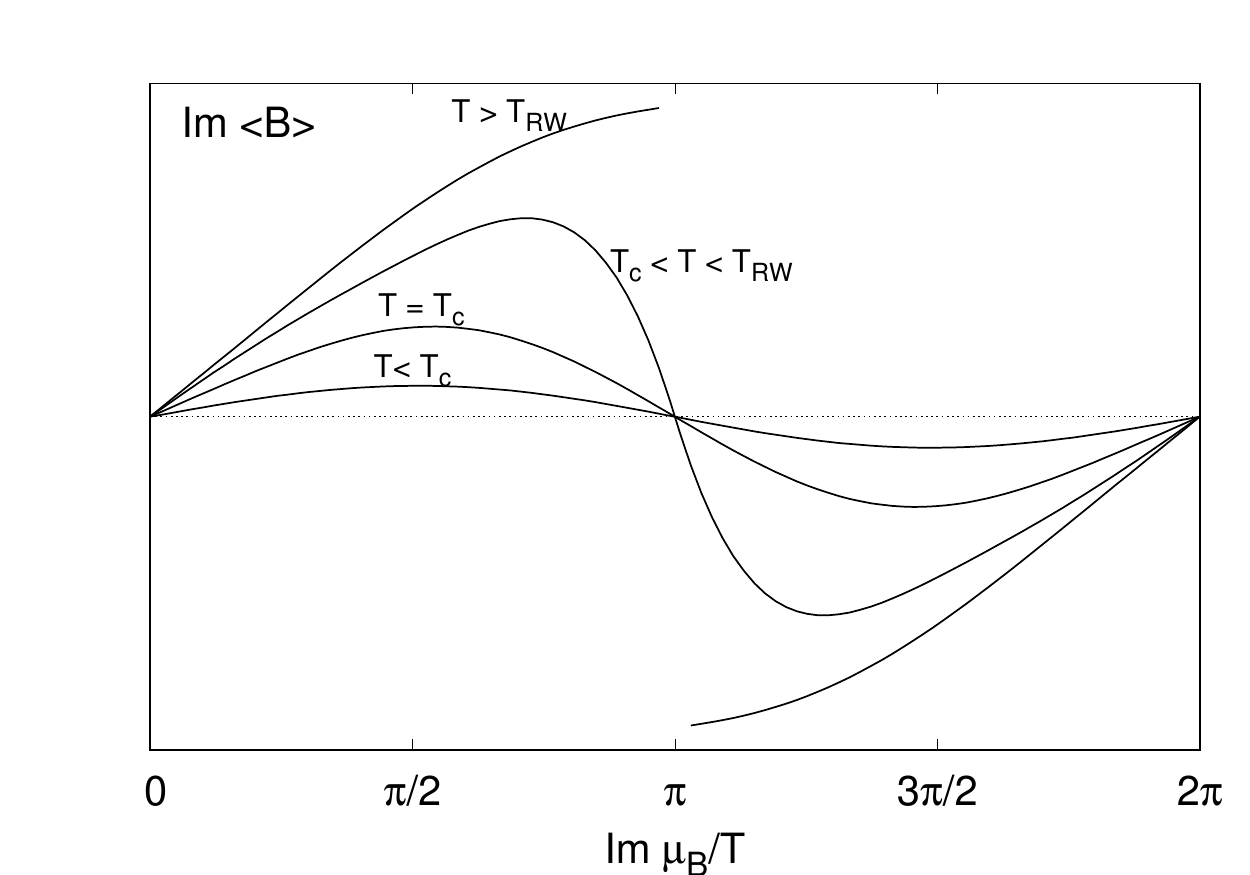}
\end{center}
\caption{\label{fig:imnq_illustration}
Cartoon for the imaginary baryon number ($\mathrm{Im}~\chi^B_1$) as a function of the imaginary chemical potential.
$T_{RW}$ is the temperature of the Roberge-Weiss critical point.
}
\end{figure}

At $\mu_B= i \pi T$ there is a first order
phase transition at all temperatures above the Roberge-Weiss critical end point
$T_{RW}$ \cite{Roberge:1986mm}. When $\mu_B$ crosses $i\pi T$ in the imaginary
direction, the imaginary baryon density is discontinuous. 
This behaviour is illustrated in Fig.~\ref{fig:imnq_illustration},
where the imaginary baryon density as a function of the imaginary chemical potential
is shown. At low temperature the Hadron Resonance Gas model predicts 
$\langle B\rangle \sim \sinh (\mu_B/T)$, thus for imaginary values we expect a sine function below $T_c$:
$\mathrm{Im}\langle B\rangle \sim \sin(\mathrm{Im}\mu_B/T)$.
At temperatures slightly above $T_c$, we observe that further Fourier
components appear in addition to $\sin(\mathrm{Im}\mu_B/T)$
with alternating coefficients, these are consistent with a repulsive interaction
between baryons \cite{Vovchenko:2017xad}.
At very high temperatures, on the other hand, $\langle B\rangle$ is a polynomial of $\mu_B$
since the diagrams contributing to its $\sim\mu_B^5$ and higher order components are suppressed by
asymptotic freedom \cite{kapusta:book,Vuorinen:2002ue}.
The Stefan-Boltzmann limit is non-vanishing only for two Taylor coefficients
of $\mathrm{Im}~\langle B\rangle$, giving 
$\left. \mathrm{Im} \langle B\rangle\right|_{\mu_B/T=i\pi-\epsilon}=8\pi/27$.
At finite temperatures above $T_{RW}$ this expectation value is smaller but positive.
By Eq.~(\ref{eq:RWconj}), it implies a first order transition at $\mu_B=i\pi T$.

The order of the transition at $T_{RW}$ heavily depends on the quark masses
\cite{Philipsen:2014rpa,Cuteri:2015qkq}.  For physical quark masses one
obtains $T_{RW}=208(5)~$MeV, and the scaling around the end-point is
consistent with the Ising exponents \cite{Bonati:2016pwz}.  This implies that,
for physical parameters, the transition is limited to $\mu_B=i\pi T$ without any
other structures between the imaginary interval $[0,i\pi)$
\cite{Philipsen:2014rpa}.

Thus, we have only the range $\mu/T \in [0,i\pi)$ to explore the
$\mu$-dependence of the observables.  Recent simulations in this range include
the determination of the transition line, where the slope was determined on the
negative side of  the $T - \mu_B^2$ phase diagram. Using analyticity arguments,
this coefficient gives the curvature of the transition line on the real $T
-\mu_B$ phase diagram \cite{Bonati:2015bha,Bellwied:2015rza,Cea:2015cya}.
Apart from the transition temperature, we used imaginary chemical potentials
also to extrapolate the equation of state to real $\mu_B$
\cite{Gunther:2016vcp}, which serves as an alternative approach to the Taylor
extrapolation \cite{Bazavov:2017dus}. In an recent study D'Elia et al. have
used the low order fluctuations at imaginary chemical potentials to
calculate generalized quark number susceptibilities \cite{DElia:2016jqh}.

\section{Analysis details\label{sec:ana}}

\subsection{Lattice setup\label{sec:setup}}

In this work we calculate high order fluctuations by studying the imaginary
chemical potential dependence of various generalized quark number susceptibilities.

We use a tree-level Symanzik improved gauge action, with four times stout
smeared ($\rho = 0.125$) staggered fermions. We simulate $2+1+1$ dynamical
quarks, where the light flavors are tuned in a way to reproduce the physical
pion and kaon masses and we set $\frac{m_c}{m_s} = 11.85$
\cite{McNeile:2010ji}.  For the zero-temperature runs that we used for the determination of
the bare masses and the coupling, the volumes satisfy $L m_{\pi} > 4$. 
The scale is determined via $f_{\pi}$. More details on the scale setting and
lattice setup can be found in \cite{Bellwied:2015lba}.

Our lattice ensembles are generated at eighteen temperatures
in the temperature range 135\dots 220~MeV. 
We simulate at eight different values of imaginary $\mu_B$ given as: $
\mu_B^{(j)} = i T \frac{j \pi}{8}$ for $j \in \{0, 1, 2, 3, 4, 5,6,7\}$.
In this work the analysis is done purely on a $48^3 \times 12$ lattice, we leave the continuum
extrapolation for future work. 

In terms of quark chemical potentials we generate ensembles with $\mu_u=\mu_d=\mu_s=\mu_B/3$. 
In each simulation point we calculate all derivatives in Eq.~(\ref{eq:chideriv}) up to fourth order.
Thanks to our scan in $\mathrm{Im}~\hat\mu_B$, we can calculate additional $\mu_B$ derivatives.
Ref.~\cite{DElia:2016jqh} uses various ``trajectories'' in the $\mu_B-\mu_Q-\mu_S$ space, allowing
the numerical determination of higher e.g. $\mu_Q$ and $\mu_S$ derivatives. We find relatively good signal
for the $\mu_Q$ and $\mu_S$ derivatives by directly evaluating Eq.~(\ref{eq:chideriv}) within one simulation.
We recently summarized the details of the direct calculation in Ref.~\cite{Bellwied:2015lba}.

\subsection{Correlated fit with priors\label{sec:fit}}

We start with the analysis for $\chi_2^B(T)$, $\chi_4^B(T)$ and $\chi_6^B(T)$. Our goal is to
calculate these quantities at zero chemical potential, using the imaginary chemical potential data up to
$\chi_B^4(T,\hat\mu_B)$. In this work we extract these derivatives at a fixed temperature. Results
for different temperatures are obtained completely independently, an interpolation in temperature
is not necessary at any point. Thus, the error bars in our results plot will be independent.
The errors between the quantities $\chi_2^B(T)$, $\chi_4^B(T)$ and $\chi_6^B(T)$ will be highly correlated, though,
since these are extracted through the same set of ensembles at the given
temperature. This correlation will be taken into account when combined
quantities are calculated, or when an extrapolation to real chemical potential
is undertaken.

Thus we consider the ensembles at a fixed temperature $T$. For each value of
imaginary $\mu_B \neq 0$ we determine $\chi_1^B$, $\chi_2^B$, $\chi_3^B$ and
$\chi_4^B$ from simulation, while for $\mu_B=0$ only $\chi_2^B$ and $\chi_4^B$
can be used, since $\chi_1^B$ and $\chi_3^B$ are odd functions of $\mu_B$
and therefore equal to zero. 

We make the ansatz for the pressure:
\begin{align}
 \chi_0^B(\hat \mu_B) &= c_0 + c_2  \hat \mu_B^2 + c_4  \hat \mu_B^4 +c_6  \hat \mu_B^6 + c_8  \hat \mu_B^8+ c_{10}  \hat \mu_B^{10},
\end{align}
where the Taylor expansion coefficients $c_n$ are related to the baryon number fluctuations $\chi_n^B$ by: $n!c_n=\chi_n^B$. Our data do not allow for an independent determination of $c_8$ and $c_{10}$. Nevertheless, in order to have some control over these terms we make the assumption
\begin{align}
 |\chi^B_8| &\lesssim \chi_4^B\\
 |\chi^B_{10}| &\lesssim \chi_4^B
\label{eq:chi_constraint}
\end{align}
or in terms of the $c_n$ coefficients
\begin{align*}
 8! c_8 \lesssim 4!c_4\\
 10! c_{10} \lesssim 4!c_4.
\end{align*}

We can then rewrite our ansatz as
\begin{align}
 \chi_0^B(\hat \mu_B) &= c_0 + c_2  \hat \mu_B^2 + c_4  \hat \mu_B^4 +c_6  \hat \mu_B^6 + \frac{4!}{8!} c_4 \epsilon_1  \hat \mu_B^8+ \frac{4!}{10!} c_4 \epsilon_2  \hat \mu_B^{10} .
 \label{deriv}
\end{align}
where $\epsilon_1$ and $\epsilon_2$ are drawn randomly from a normal distribution with mean -1.25 and variance 2.75. We use the same distribution for all temperatures.

In effect, our $c_8$ and $c_{10}$ coefficients are stochastic variables. There is sufficient probability to obtain coefficients
that slightly break Eq.~(\ref{eq:chi_constraint}), should the data prefer larger than expected fluctuations, e.g. due to
a nearby critical end point. The used distribution for $\epsilon_{1,2}$ actually implements a prior for $\chi^B_8$ and
$\chi^B_{10}$. 
In the HRG model we know that
$\chi^B_2=\chi^B_4=\chi^B_6=\chi^B_8=\chi^B_{10}$, which is well represented by
the prior distribution.  At high temperatures $\chi^6_B$ and higher
coefficients quickly approach zero, as obtained in Hard Thermal
Loop results \cite{Haque:2013sja}. In the transition regime, the higher moments of the baryon fluctuations
are dominated by the fact that the transition line is $\mu_B$-dependent. Starting from $\mu_B=0$ at a fixed
temperature between $T_c$ and $T_{RW}$, a crossover line is developed as the imaginary chemical potential is
introduced. The magnitude of higher order fluctuations in the transition regime can be estimated by a very
simple observation. The behaviour of the quark density $\chi^B_1(T,\hat\mu_B)$ is reasonably approximated
by $\mu_B \chi^B_2 ( T + T_c \kappa\hat\mu_B^2,\hat\mu_B=0)$, 
where $\kappa$ is the curvature of the transition line in the $\mu_B-T$ phase diagram. In this approximation,
the only source of $\mu_B$-dependence is coming from the curvature of the transition line.
Calculating the $\mu_B$ derivatives gives a basic estimate for $\chi^B_8$,
which we used to tune the prior distribution.

For this ansatz we calculate the following derivatives, which are the actually simulated lattice observables:
\begin{align}
 \chi_1^B(\hat \mu_B) &= 2 c_2  \hat \mu_B + 4c_4  \hat \mu_B^3 +6c_6  \hat \mu_B^5+ \frac{4!}{7!} c_4 \epsilon_1  \hat \mu_B^7+ \frac{4!}{9!} c_4 \epsilon_2  \hat \mu_B^{9}  \label{eq:ansatz_chib1} \\
\chi_2^B(\hat \mu_B) &=2 c_2   + 12c_4  \hat \mu_B^2 + 30 c_6  \hat \mu_B^4+ \frac{4!}{6!} c_4 \epsilon_1  \hat \mu_B^6+ \frac{4!}{8!} c_4 \epsilon_2  \hat \mu_B^{8}  \\
\chi_3^B(\hat \mu_B) &=  24 c_4  \hat \mu_B + 120 c_6  \hat \mu_B^3+ \frac{4!}{5!} c_4 \epsilon_1  \hat \mu_B^5+ \frac{4!}{7!} c_4 \epsilon_2  \hat \mu_B^{7}  \\
 \chi_4^B(\hat \mu_B) &= 24 c_4   + 360 c_6  \hat \mu_B^2+  c_4 \epsilon_1  \hat \mu_B^4+ \frac{4!}{6!} c_4 \epsilon_2  \hat \mu_B^{6} . \label{eq:ansatz_chib4}
\end{align}
We perform a correlated fit for the four measured observables, thus obtaining
the values of $c_2$,  $c_4$ and $c_6$ for each temperature, and the
corresponding $\chi_2^B$, $\chi_4^B$ and $\chi_6^B$. 
We repeat the fit for 1000 random draws for $\epsilon_1$ and $\epsilon_2$. The result is weighted using
the Akaike Information Criterion \cite{Akaike}. Through these weights we get a posterior
distribution from the prior distribution. Our final estimate for $\chi_8^B$ represents this
posterior distribution. We do not show the posterior for $\chi_{10}^B$, which is mostly noise.

\begin{figure}
 \begin{center}
 \includegraphics[width = \textwidth]{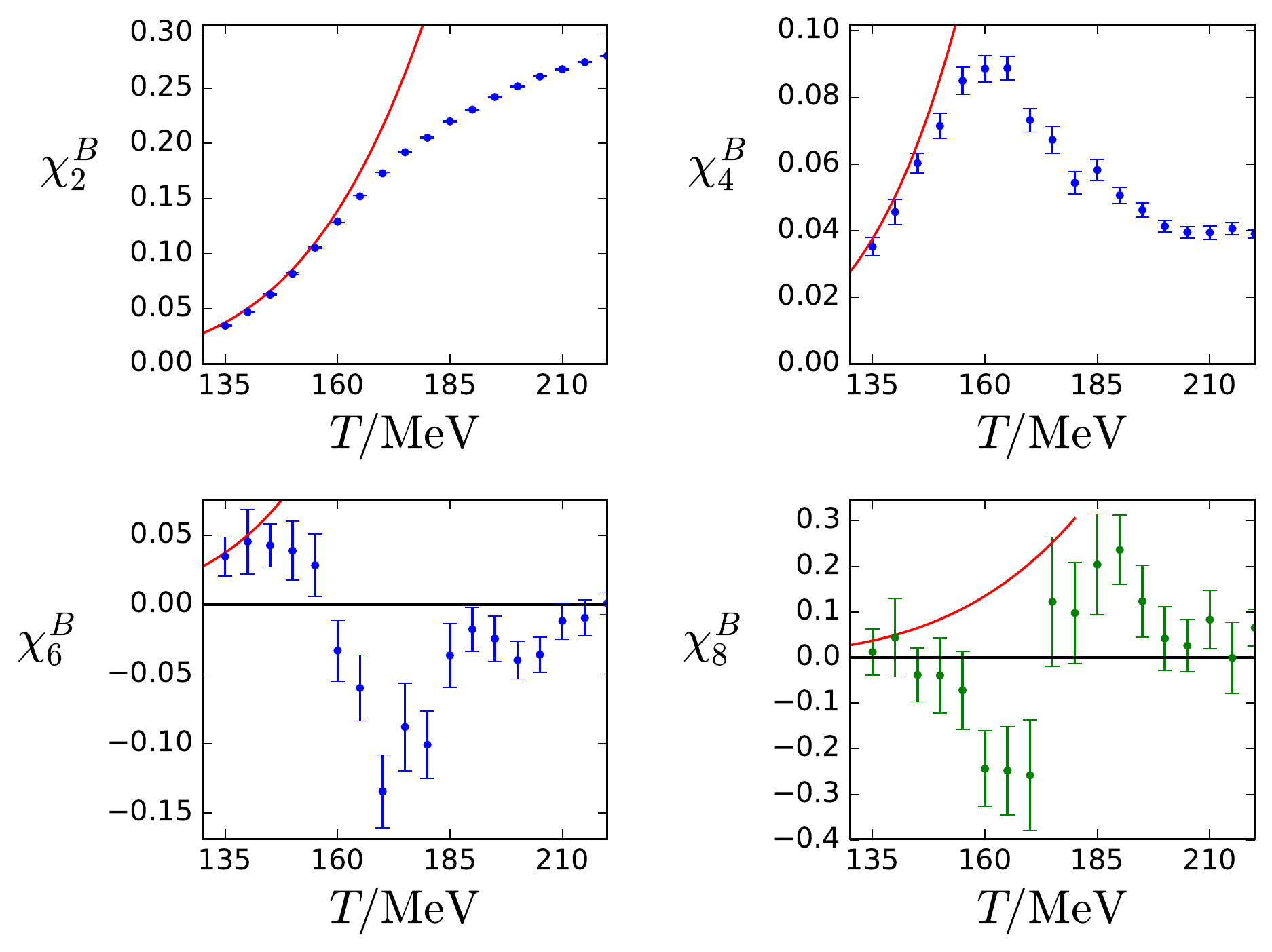}
 \end{center}
 \caption{Results for $\chi^B_2$, $\chi^B_4$, $\chi^B_6$ and an estimate for
$\chi_8^B$ as functions of the temperature, obtained from the
single-temperature analysis. We plot $\chi^B_8$ in green to point out that its
determination is guided by a prior, which is linked to the $\chi^B_4$ observable by Eq.
(\ref{deriv}). The red curve in each panel corresponds to the Hadron Resonance
Gas (HRG) model result. 
\label{fig:X6}} \end{figure}

These results are shown in Fig. \ref{fig:X6}, together with an estimate of $\chi_8^B$, related to
$\chi_4^B$ by Eq. (\ref{deriv}).

\subsection{Cross-correlators\label{sec:cross}}

So far we only considered  derivatives with respect to the baryonic chemical potential.
In our previous, direct analysis in Ref.~\cite{Bellwied:2015lba},
the $\mu_B$-derivatives had larger errors than $\mu_Q-$ or $\mu_S-$derivatives. For $\mu_Q$, the most noisy disconnected contributions come with
smaller prefactors, while for $\mu_S$ the disconnected contributions are
small due to the heavier strange mass. Our approach was designed to
improve the $\mu_B$-derivatives only. 
Therefore, the $\mu_S$ and $\mu_Q$ derivatives have to be simulated directly
and without the support from the fit that we used in the $\mu_B$ direction. 
Our result on $\chi^{QS}_{jk}$ improved only due to the increase in the
statistics since \cite{Bellwied:2015lba}. 

On the other hand, baryon-strange and baryon-charge mixed derivatives do benefit from the
imaginary $\mu_B$ data.
We simulate various $ \chi^{B,Q,S}_{i,j,k}$ with the appropriate values of $j$ and
$k$ and all possible values of $i$ so that $ i+j+k \leq 4$.
For each group of fluctuations with the same $j$ and $k$ we perform a fit
analogous to the procedure described in Section~\ref{sec:fit}.

Let's take the example of $j=1, k=0$.
Our ansatz for cross-correlators is analogous to
Eqs.~(\ref{eq:ansatz_chib1})-(\ref{eq:ansatz_chib4}):

\begin{align}
\chi_{01}^{BS}(\hat \mu_B) =
\chi_{11}^{BS} \hat\mu_B
+ \frac{1}{3!} \chi_{31}^{BS} \hat\mu_B^3
+ \frac{1}{5!} \chi_{51}^{BS} \hat\mu_B^5
+ \frac{1}{7!} \chi_{71}^{BS} \hat\mu_B^7
+ \frac{1}{9!} \chi_{91}^{BS} \hat\mu_B^9
\label{eq:ansatz_S}
\end{align}

We truncated the expression at tenth order. The priors assume $|\chi_{71}^{BS}|\lesssim |\chi_{31}^{BS}|$
and $|\chi_{91}^{BS}|\lesssim |\chi_{31}^{BS}|$, as it is certainly true at high temperature and within the HRG model.
The prior distribution is wider than 1, we used the same mean and variance as in the channel
with no $\mu_S$ derivative.

\begin{figure}[h!]
 \begin{center}
  \includegraphics[width = 0.8\textwidth]{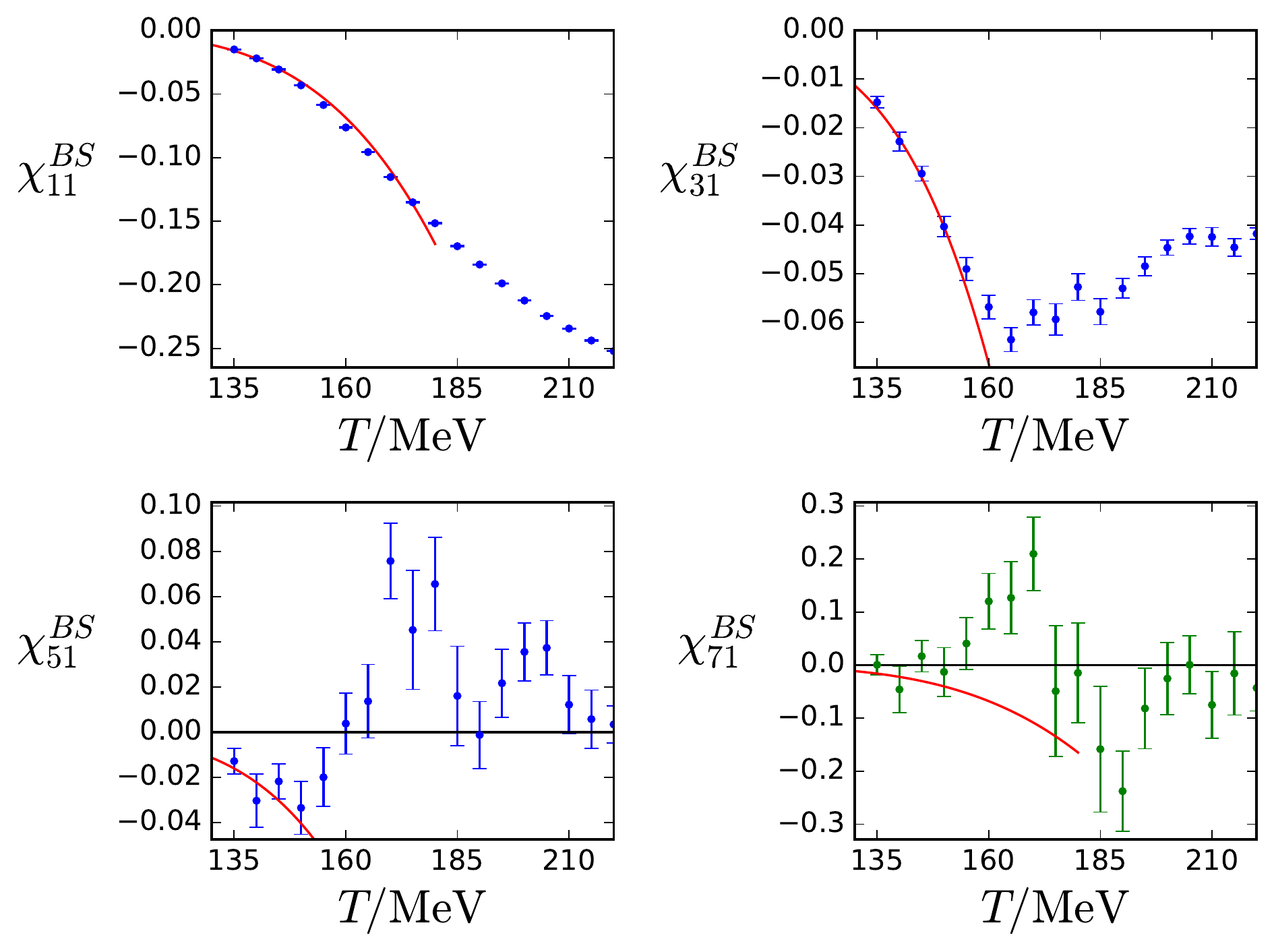}
 \end{center}
 \caption{$\chi_{11}^{BS},~\chi_{31}^{BS},~\chi_{51}^{BS}$ and an estimate for $\chi_{71}^{BS}$ as functions of the temperature. The red curves are the HRG model results. 
\label{fig:resultS}
}
\end{figure}

When we use Eq.~(\ref{eq:ansatz_S}) we take 
$\chi^{S}_{1}$, $\chi^{BS}_{11}$, $\chi^{BS}_{21}$ and $\chi^{BS}_{31}$ as correlated quartets for each
imaginary chemical potential and determine the three free coefficients of Eq.~(\ref{eq:ansatz_S}).
This fitting procedure is repeated 1000 times with random
$\chi^{BS}_{71}/\chi^{BS}_{31}$ and $\chi^{BS}_{91}/\chi^{BS}_{31}$ coefficients.
Again, using the Akaike weights we constrain the prior distribution. The resulting estimate for $\chi_{71}^{BS}$
along with the fit coefficients are shown in Fig.~\ref{fig:resultS}. The posterior for $\chi_{91}^{BS}$ is not
only noisy, but it is probably heavily contaminated by the higher orders that we did not account for.

The other channels with higher $\mu_S$ or $\mu_Q$ derivatives are obtained analogously.
These are plotted in Appendix~\ref{sec:correlators}.

\subsection{Error Analysis\label{sec:error}}
For a reliable comparison between experimental measurements and theoretical
calculations, the error estimate is an important ingredient. Our statistical
error is estimated through the jackknife method. For our systematic error
there are several sources. We determine our systematic error by the histogram
method described in \cite{Durr:2008zz}, where each analysis is weighted with
the Akaike information criteria.  We include the influence of the number of
points in the $\mu_B$ direction, by either including or ignoring the data from
our highest value of $\mu_B$. A very important source for our systematic error is the
influence of the higher order contributions in $\mu_B$. This effect was estimated
by adding the higher order terms with pre-factors $\epsilon_1$ and $\epsilon_2$
as described in Section \ref{sec:fit}. We consider 1000 different $\epsilon$ pairs and
add the different analyses to our histogram. The width of the histogram
using Akaike weights corresponding to the fit quality gives the systematic errors
for the fit coefficients, and from the same histogram we obtain the posterior 
distributions for $\epsilon_1$. The physical quantities that are constrained only by
the posterior distribution are plotted with green symbols.

These histograms are built independently for each number ($j$ and $k$) of $\mu_S$ and $\mu_Q$ derivatives.
When calculating the systematics for the cumulant ratios (Section \ref{sec:cumulants}) we
need to calculate different combinations of diagonal and non-diagonal fluctuations
from the available analyses. Though these fits (corresponding to the same temperature) are carried out
separately we keep track of the statistical correlation, by maintaining the jackknife ensembles
throughout the analysis. The correct propagation of systematic errors is a more elaborate
procedure. When $\chi^{BSQ}_{ijk}$ coefficients are combined with different $j,k$ pairs,
different histograms have to be combined. If we had only two variables to combine,
each of the 2000 first fit variants should be combined with each of the 2000
second fit variants and use the product of the respective probability weights.
Instead, we combine the fit results by drawing 'good' fits by importance
sampling from each histogram independently. In this way, $\mathcal{O}(100)$
random combinations of $\chi^{BSQ}_{ijk}$ results already give convergence for
each discussed quantity and its error bar. For the results in this paper
we used 1000 such random combinations.  This procedure assumes that between
different ${j,k}$ pairs the prior distribution is uncorrelated.

\section{Phenomenology at finite chemical potential\label{sec:cumulants}}

For a comparison with heavy ion collision experiments, the cumulants of the
net-baryon distribution are very useful observables. The first four cumulants
are the mean $M_B$, the variance $\sigma_B^2$, the skewness $S_B$ and the
kurtosis $\kappa_B$. By forming appropriate ratios, we can cancel out explicit
volume factors. However, the measured distributions themselves may still depend
on the volume, which one should take into account when comparing to
experiments. 

Heavy ion collisions involving lead or gold atoms at $\mu_B > 0$ correspond to the following situation
\begin{equation}
\langle n_S \rangle = 0 \, \qquad \langle n_Q\rangle = 0.4 \langle n_B\rangle\,.
\label{eq:neutrality}
\end{equation}
For each $T$ and $\mu_B$ pair, we have to first calculate $\mu_Q$ and $\mu_S$ that satisfy
this condition. The resulting $\mu_Q(\mu_B)$ and $\mu_S(\mu_B)$ functions, too,
can be Taylor expandend \cite{Bazavov:2012vg,Borsanyi:2013hza}, introducing
\begin{align}
q_j&=\left.\frac{1}{j!} \frac{d^j\hat\mu_Q}{(d\hat\mu_B)^j}\right|_{\mu_B=0}\\
s_j&=\left.\frac{1}{j!} \frac{d^j\hat\mu_S}{(d\hat\mu_B)^j}\right|_{\mu_B=0}.
\end{align}

We investigate three different ratios of cumulants:
\begin{equation}
  \frac{M_B}{\sigma^2_B} = \frac{\chi^B_1(T,\hat \mu_B)}{\chi^B_2(T,\hat \mu_B)}= \hat \mu_B r_{12}^{B,1} + \hat \mu_B^3 r_{12}^{B,3} + \ldots \label{eq:1}
 \end{equation}
 
 \begin{equation}
  \frac{S_B \sigma_B^3}{M_B} = \frac{\chi^B_3(T,\hat \mu_B)}{\chi^B_1(T,\hat \mu_B)}= r_{31}^{B,0} + \hat \mu_B^2 r_{31}^{B,2} + \ldots \label{eq:2}
 \end{equation}
 
 \begin{equation}
  \kappa_B  \sigma^2_B= \frac{\chi^B_4(T,\hat \mu_B)}{\chi^B_2(T,\hat \mu_B)}= r_{42}^{B,0} + \hat \mu_B^2 r_{42}^{B,2} +  \hat \mu_B^4 r_{42}^{B,4} + \ldots
\label{eq:3}
 \end{equation}
 
The $\mu_B$-dependence of the $\chi^B_i(T,\hat \mu_B)$ can again be written as a Taylor series:
 \begin{align}
\chi^{BQS}_{i,j,k} (\hat\mu_B) &=
\chi^{BQS}_{i,j,k} (0) 
+ \hat\mu_B \left [\chi^{BQS}_{i+1,j,k}(0) 
+ q_1\chi^{BQS}_{i,j+1,k}(0) 
+ s_1 \chi^{BQS}_{i,j,k+1}(0)  \right]
\nonumber\\
&+
\frac12 \hat\mu_B^2 \left [
  \chi^{BQS}_{i+2,j,k}(0) 
  + q_1^2 \chi^{BQS}_{i,j+2,k}(0) 
  + s_1^2 \chi^{BQS}_{i,j,k+2}(0)  \right.
  \nonumber\\
&
\left.
  + 2 q_1 s_1 \chi^{BQS}_{i,j+1,k+1}(0) 
  + 2 q_1 \chi^{BQS}_{i+1,j+1,k}(0) 
  + 2 s_1 \chi^{BQS}_{i+1,j,k+1}(0) 
\right]+\dots\,.
\label{eq:4}
\end{align}

The $\chi$ coefficients that we determined in Section \ref{sec:ana} include derivatives
up to sixth order, and we have estimates for the eighth order, too. The fit coefficients 
corresponding to the tenth order are likely to be contaminated by higher orders, that
we did not include into the ansatz. These $\chi^{BQS}_{ijk}$ coefficients, however,
are given for $j+k\le4$, which is the highest order that we used in  $\mu_Q$ and $\mu_S$.

\begin{figure}
 \begin{center}
  \includegraphics[width = \textwidth]{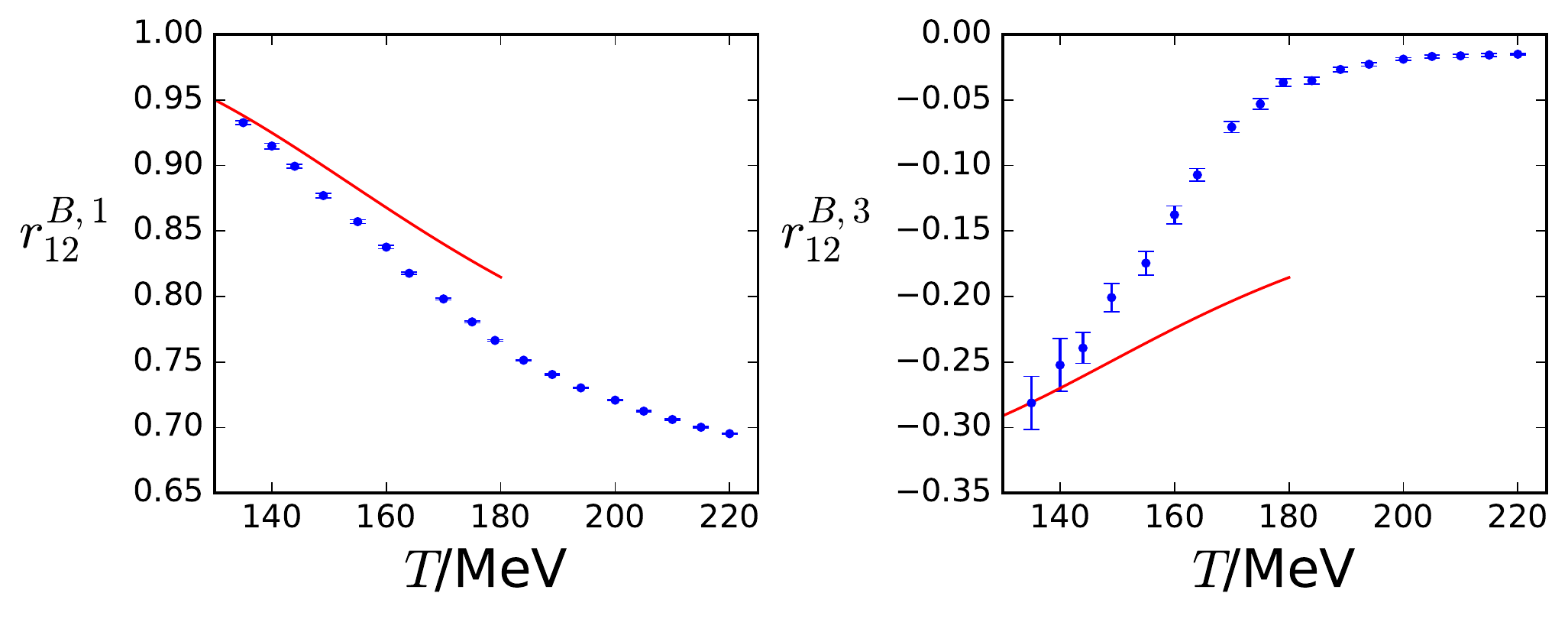}
 \end{center}
 \caption{Taylor expansion coefficients for $\frac{M_B}{\sigma^2_B}= \frac{\chi^B_1(T,\hat \mu_B)}{\chi^B_2(T,\hat \mu_B)}$ as functions of the temperature: $ r_{12}^{B,1}$ (left panel) and $ r_{12}^{B,3}$ (right panel). \label{fig:r1}} 
\end{figure}

\begin{figure}
 \begin{center}
  \includegraphics[width = \textwidth]{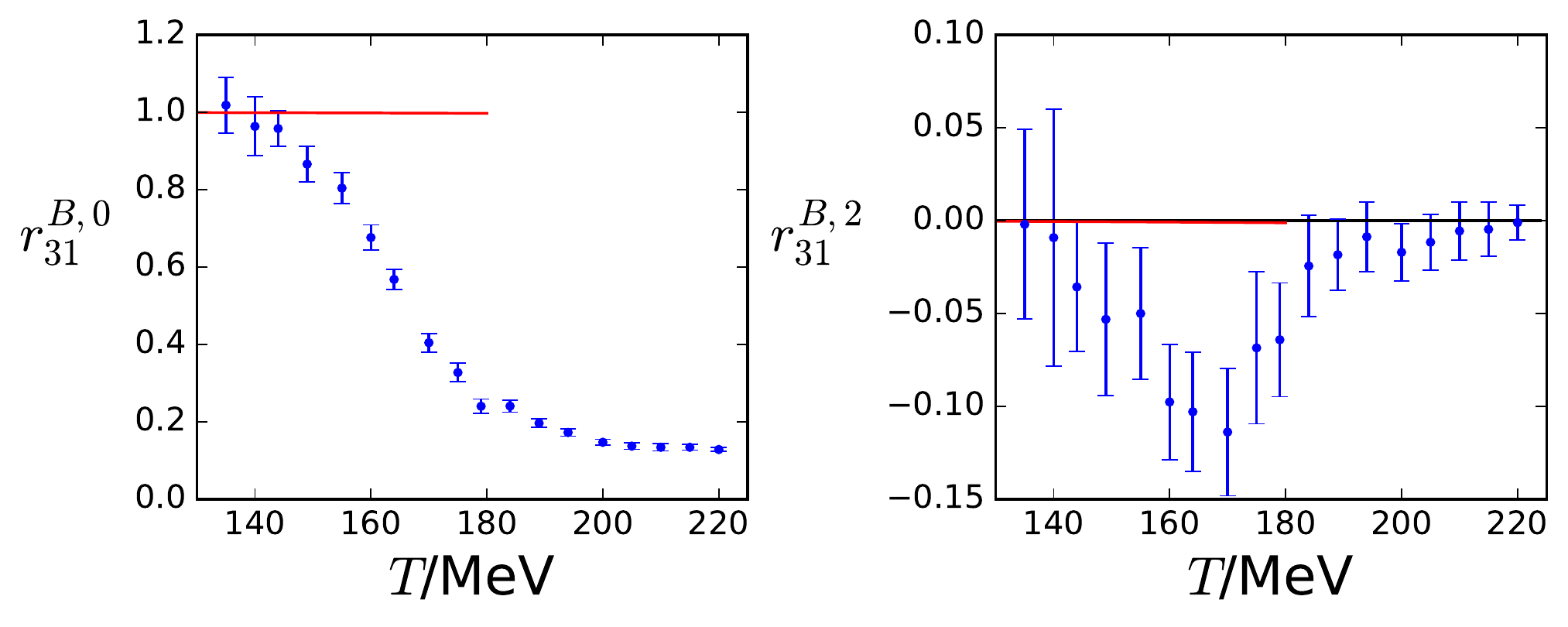}
 \end{center}
 \caption{Taylor expansion coefficients for $\frac{S_B \sigma_B^3}{M_B} = \frac{\chi^B_3(T,\hat \mu_B)}{\chi^B_1(T,\hat \mu_B)}$ as functions of the temperature: $r_{31}^{B,0}$ (left panel) and $r_{31}^{B,2}$ (right panel). \label{fig:r2}}
\end{figure}

\begin{figure}
 \begin{center}
  \includegraphics[width = \textwidth]{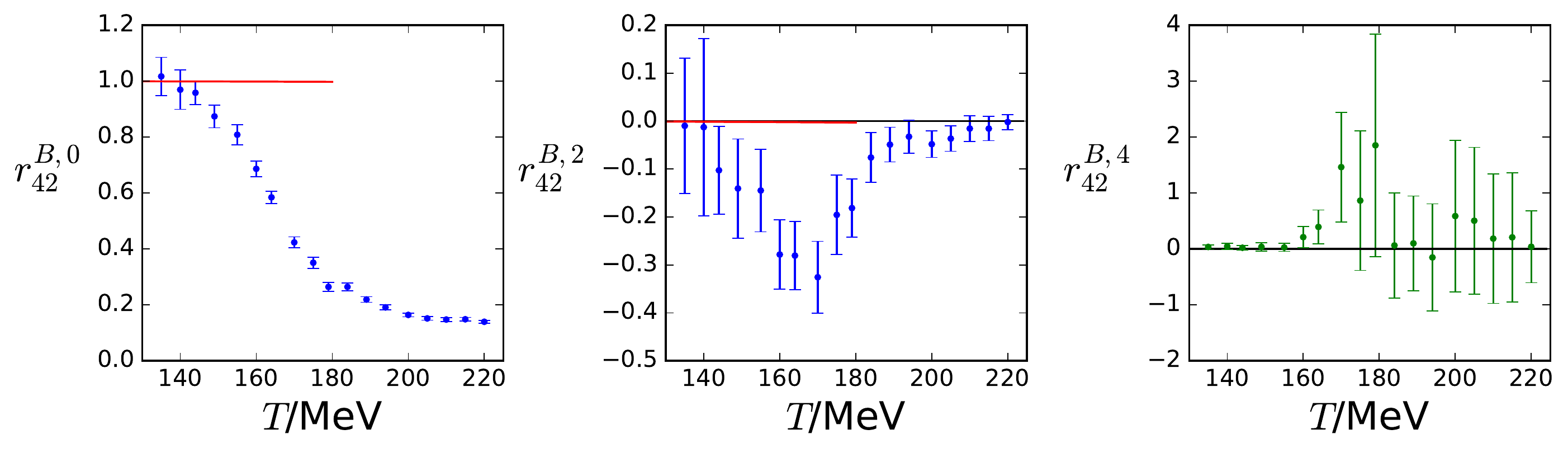}
 \end{center}
 \caption{Taylor expansion coefficients for $\kappa_B  \sigma^2_B= \frac{\chi^B_4(T,\hat \mu_B)}{\chi^B_2(T,\hat \mu_B)}$ as functions of the temperature: $r_{42}^{B,0}$ (left panel) $ r_{42}^{B,2}$ (middle panel), $ r_{42}^{B,4}$ (right panel). The latter is not obtained independently, but by means of the prior ansatz (see text): for this reason, we plot it in green. \label{fig:r3}}
\end{figure}

This list of coefficients allows us to calculate the $r^{B,k}_{ij}$
coefficients from Equations (\ref{eq:1}), (\ref{eq:2}) and (\ref{eq:3}). 
The results for the  $r^{B,k}_{ij}$ coefficients are shown in Figures \ref{fig:r1},  \ref{fig:r2} and \ref{fig:r3}. 
We confirm the observation from Ref. \cite{Bazavov:2017tot} that the coefficient $r^{B,2}_{42}$
has a similar temperature dependence as $r^{B,2}_{31}$ but it is $\sim3$ times larger in magnitude.

For higher order coefficients, higher order derivatives in $\mu_S$ and $\mu_Q$
are needed. The direct simulations have a rapidly increasing error with the
order of the derivative, and very large statistics would be needed to improve
our calculations at this point. Another possibility would be to simulate new
ensembles with finite $\mu_S$ and $\mu_Q$ and do a similar fit as for the
$\mu_B$ direction. This approach has been used in \cite{DElia:2016jqh}.

\begin{figure}
 \begin{center}
 \includegraphics[width = 0.48\textwidth]{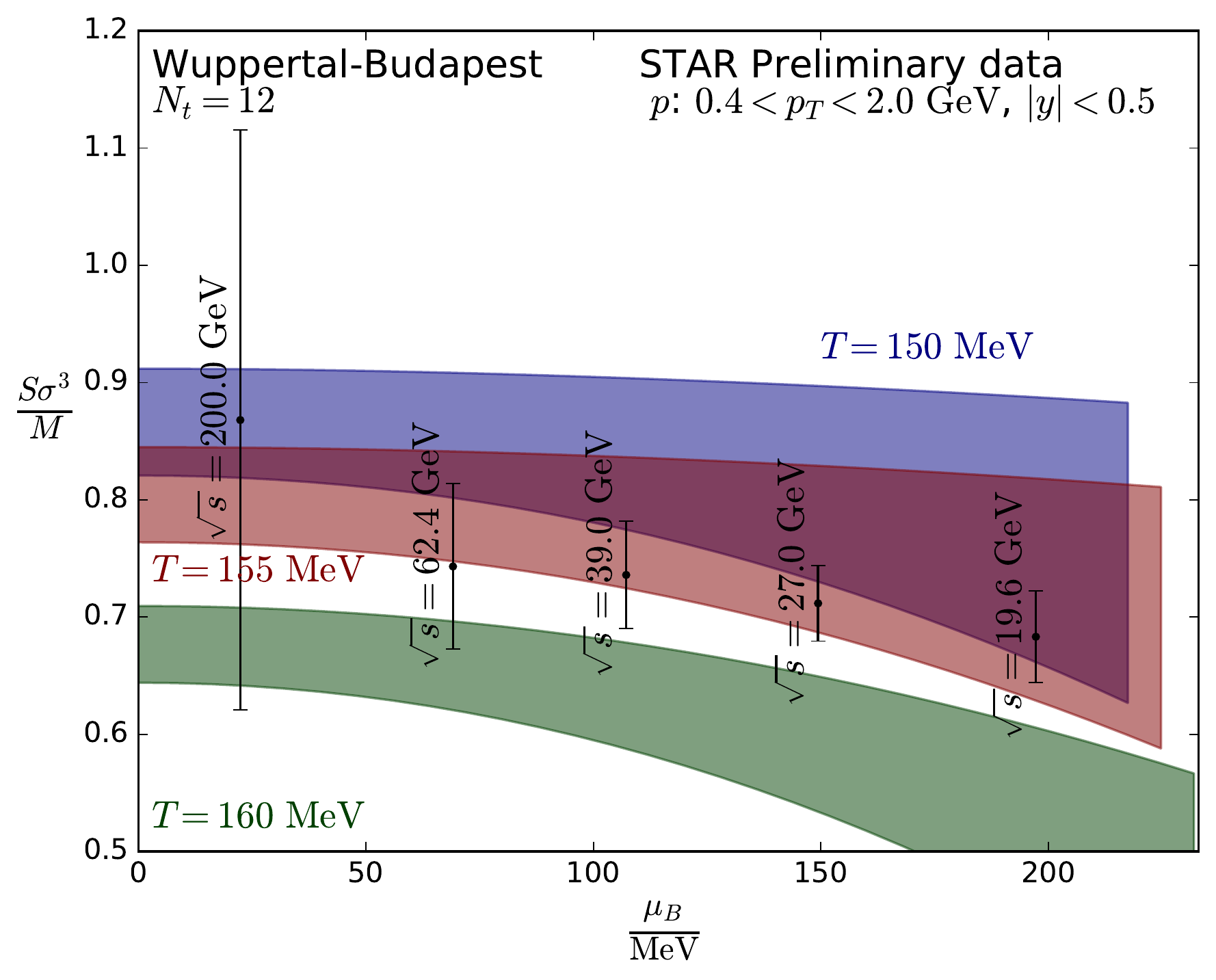}
  \includegraphics[width = 0.48\textwidth]{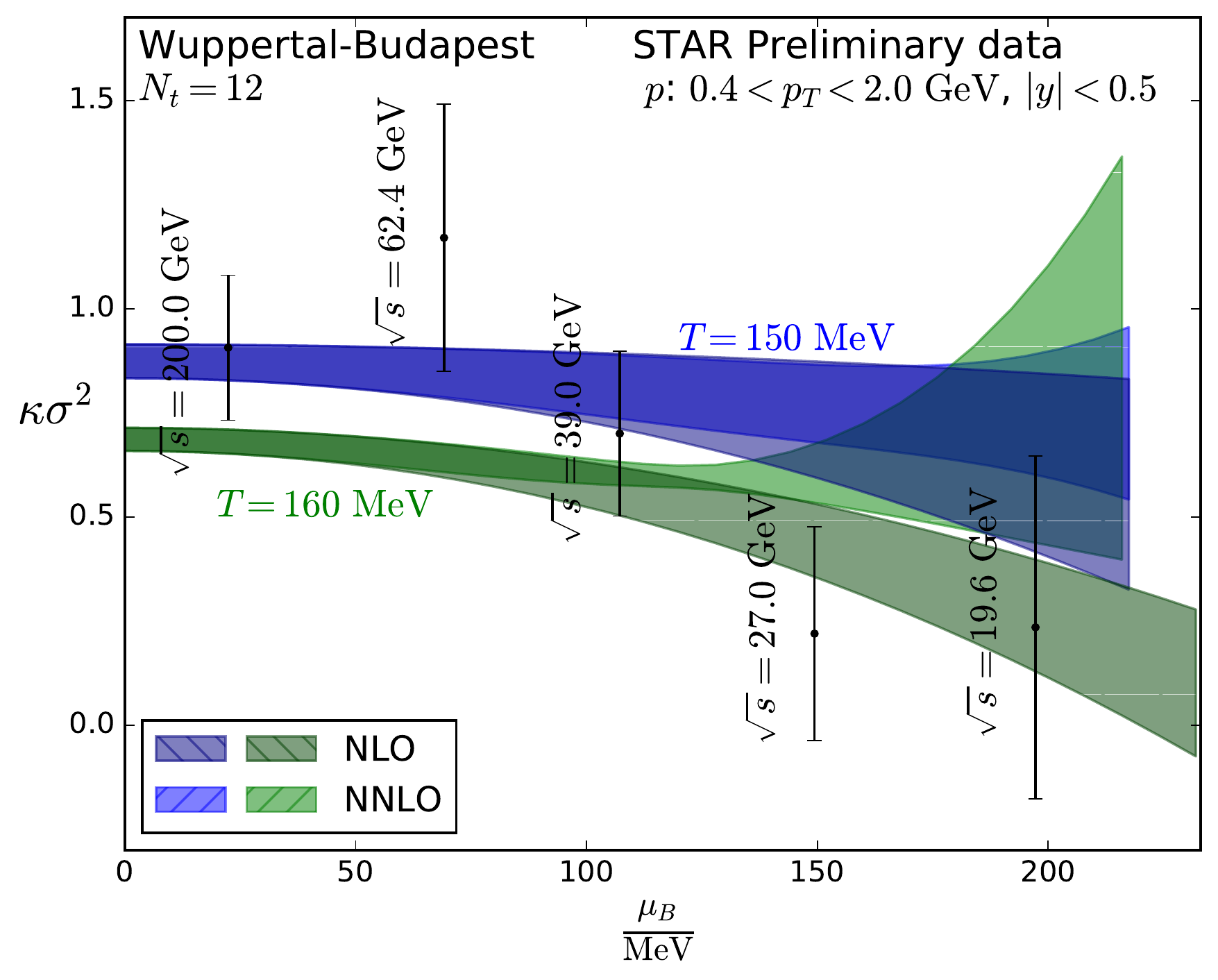}
 \end{center}
 \caption{$S_B \sigma_B^3/M_B$ (left panel) and $\kappa_B  \sigma^2_B$ (right panel) extrapolated to finite chemical potential. The left panel is extrapolated up to $\mathcal{O}(\hat{\mu}_B^2)$. In the right panel, the darker bands correspond to the extrapolation up to $\mathcal{O}(\hat{\mu}_B^2)$, whereas the lighter bands also include the $\mathcal{O}(\hat{\mu}_B^4)$ term. \label{fig:extrap}}
\end{figure}

After calculating the Taylor coefficients for $S_B \sigma_B^3/M_B$ and $\kappa_B  \sigma^2_B$, we use these results to extrapolate these quantities to finite chemical potential. They are shown in Figure \ref{fig:extrap}. In the left panel, $S_B \sigma_B^3/M_B$ is shown as a function of the chemical potential for different temperatures. The Taylor expansion for this quantity is truncated at $\mathcal{O}(\hat{\mu}_B^2)$. The black points in the figure are the experimental results from the STAR collaboration from an analysis of cumulant ratios measured at mid-rapidity, $|y|\leq0.5$, including protons and anti-protons with transverse momenta 0.4 GeV
$\leq p_t\leq2.0$ GeV \cite{Luo:2015ewa,Thader:2016gpa}. 
The beam energies were translated to chemical potentials using the fitted
formula of Ref.~\cite{Andronic:2005yp}.
Even if we do not quantitatively compare the lattice bands to the measurements to extract the freeze-out parameters, as experimental higher order fluctuations might be affected by several effects of non-thermal origin and our lattice results are not continuum extrapolated, we notice that the trend of the data with increasing $\mu_B$ can be understood in terms of our Taylor expansion.

In the right panel, we show $\kappa_B  \sigma^2_B$ as a function of $\mu_B/T$
for different temperatures. The darker bands correspond to the extrapolation up
to $\mathcal{O}(\hat{\mu}_B^2)$, whereas the lighter bands also include the
$\mathcal{O}(\hat{\mu}_B^4)$ term. Also in this case, the black points are the
experimental results from the STAR collaboration with transverse momentum cut
0.4 GeV$\leq p_t\leq2.0$ GeV \cite{Luo:2015ewa,Thader:2016gpa}. Notice that,
due to the fact that the $r_{42}^{B,4}$ is positive in the range 160 MeV$\leq
T\leq$195~MeV, we observe a non-monotonic behavior in $\kappa_B  \sigma^2_B$
for $T=160$~MeV at large chemical potentials. By comparing the two different
truncations of the Taylor series we can conclude that, as we increase the
temperature, the range of applicability of our Taylor series decreases: while
at $T=150$ MeV the two orders agree in the whole $\mu_B/T$ range shown in the
figure, at $T=160$~MeV the central line of the next-to-next-to-leading order
bends upwards and is not contained in the next-to-leading order band.
To make the NLO prediction precise substantially more computer time
would be needed.

\section{Conclusions and outlook \label{sec:con}}
In this manuscript, we have calculated several diagonal and non-diagonal fluctuations of electric charge, baryon number and strangeness up to sixth-order, in a system of 2+1+1 quark flavors with physical quark masses, on a lattice with size $48^3\times12$. The analysis has been performed simulating the lower order fluctuations at zero and imaginary chemical potential $\mu_B$, and extracting the higher order fluctuations as derivatives of the lower order ones at $\mu_B=0$. The chemical potentials for electric charge and strangeness have both been set to zero in the simulations. From these fluctuations, we have constructed ratios of baryon number cumulants as functions of $T$ and $\mu_B$, by means of a Taylor series which takes into account the experimental constraints $\langle n_S\rangle=0$ and $\langle n_Q\rangle=0.4\langle n_B\rangle$. These ratios qualitatively explain the behavior observed in the experimental measurements by the STAR collaboration as functions of the collision energy.

We focused on observables (baryon distribution, ratios of cumulants) that are less sensitive to lattice
artefacts. An obvious extension of our work will be the use of finer lattices and a continuum extrapolation.
The other extension is to use a two- or even three-dimensional mapping of the space of the imaginary chemical
potentials using non-vanishing $\mu_S$ and $\mu_Q$. That would not only improve the $\mu_S-$ and $\mu_Q-$derivatives, but would allow us to study the melting of states with
various strangeness and electric charge quantum numbers. Our first study in
this direction using strangeness chemical potentials was published in
Ref.~\cite{Alba:2017mqu}.

\section*{Acknowledgements}
This project was funded by the DFG grant SFB/TR55.
This work was supported by the Hungarian National Research,  Development and
Innovation Office, NKFIH grants KKP126769 and K113034.  An award of computer
time  was  provided  by  the  INCITE  program. This  research  used  resources
of  the  Argonne  Leadership Computing Facility,  which is a DOE Office of
Science User Facility supported under Contract DE-AC02-06CH11357.
The authors gratefully acknowledge the Gauss Centre for Supercomputing e.V.
(www.gauss-centre.eu) for funding this project by providing computing time on
the GCS Supercomputer JUQUEEN\cite{juqueen} at J\"ulich Supercomputing Centre
(JSC) as well as on HAZELHEN at HLRS Stuttgart, Germany.
This  material  is  based  upon  work  supported  by  the
National  Science  Foundation  under  grants  no. PHY-1654219 and OAC-1531814 and by the U.S. Department
of  Energy,  Office  of  Science,  Office  of  Nuclear  Physics, within the framework of the Beam Energy Scan Theory
(BEST) Topical Collaboration.  C.R. also acknowledges the support from the Center of Advanced Computing and
Data Systems at the University of Houston.
\clearpage
\appendix
\section{Results for the correlators\label{sec:correlators}}

In this Appendix we present the non-diagonal fluctuations of conserved charges
needed to construct the cumulant ratios at finite chemical potential $\mu_B$,
satisfying the constraints $\langle n_S\rangle=0$ and $\langle
n_Q\rangle=0.4\langle n_B\rangle$.

Like we did for the diagonal $\chi_i^B$, we simulate lower order fluctuations
at finite imaginary chemical potential and extract the higher order
fluctuations as derivatives of the lower order ones at $\mu_B=0$: in
particular, we simulate various $ \chi^{B,Q,S}_{i,j,k}$ with the appropriate
values of $j$ and $k$ and all possible values for $i$ so that
\begin{equation}
 i+j+k \leq 4\,,
\end{equation}
and extract the corresponding $ \chi^{B,Q,S}_{i,j,k}$ with $i+j+k \leq 6 $ and
an estimate for $i+j+k = 8 $ and sometimes even $i+j+k = 10$. By estimate
(shown in green) we mean the posterior distribution that we get for the two highest orders
when using priors, as discussed in the main text.
In total we need 15 channels to obtain all the necessary terms.

\begin{figure}[h!]
 \begin{center}
\includegraphics[width = 0.9\textwidth]{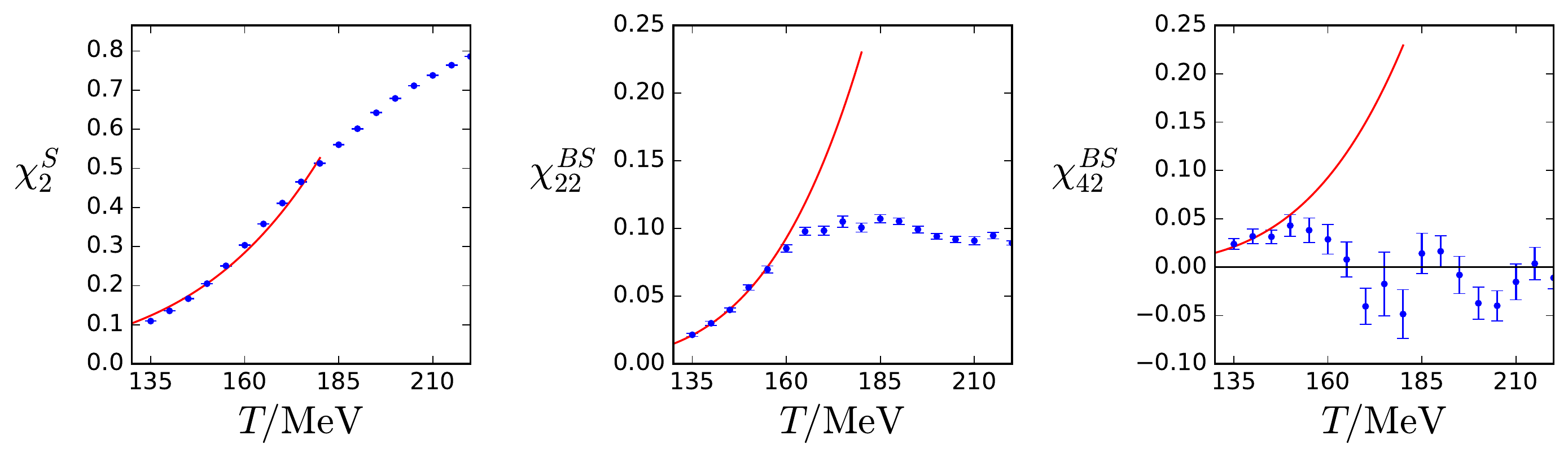}
\includegraphics[width = 0.9\textwidth]{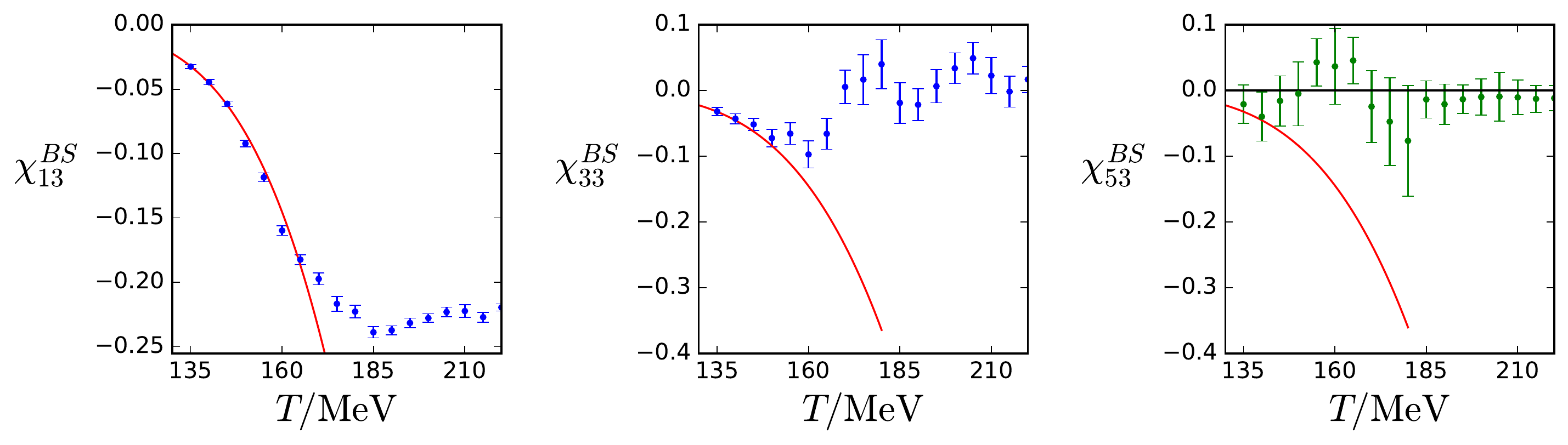}
\includegraphics[width = 0.9\textwidth]{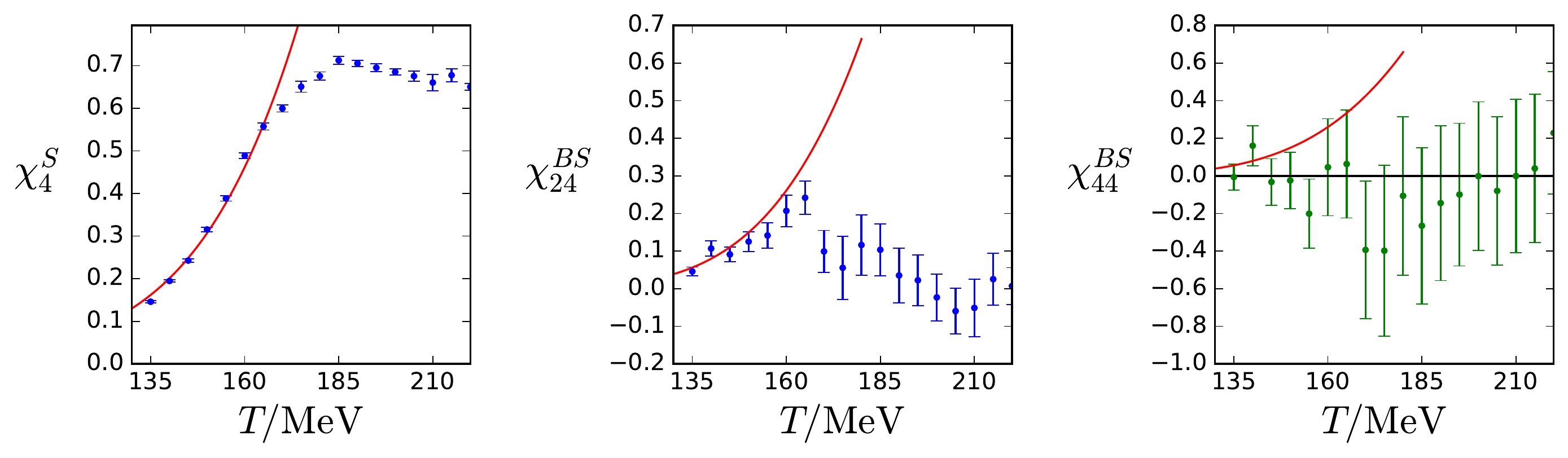}
 \end{center}
 \caption{Results containing no electric charge derivative
 on the various correlators on our $48^3\times12$ lattice as
functions of the temperature. Green data points denote our estimates for
the high orders, these were fitted using a prior distribution.
The red curves are the HRG model results.\label{resultQ0}}
\end{figure}

\begin{figure}[h!]
 \begin{center}
\includegraphics[width = 0.9\textwidth]{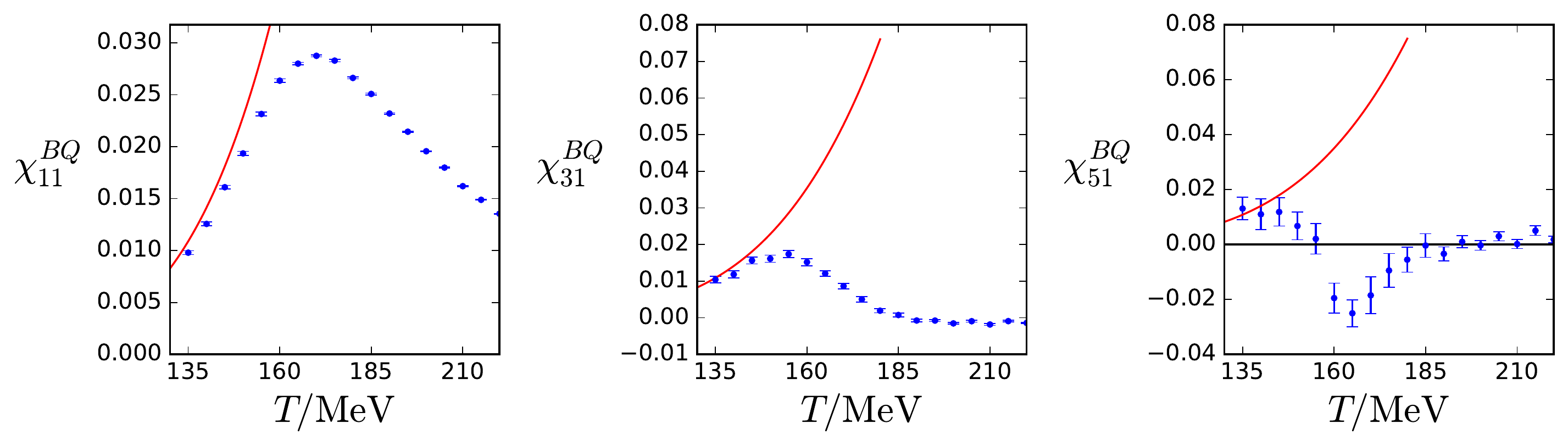}
\includegraphics[width = 0.9\textwidth]{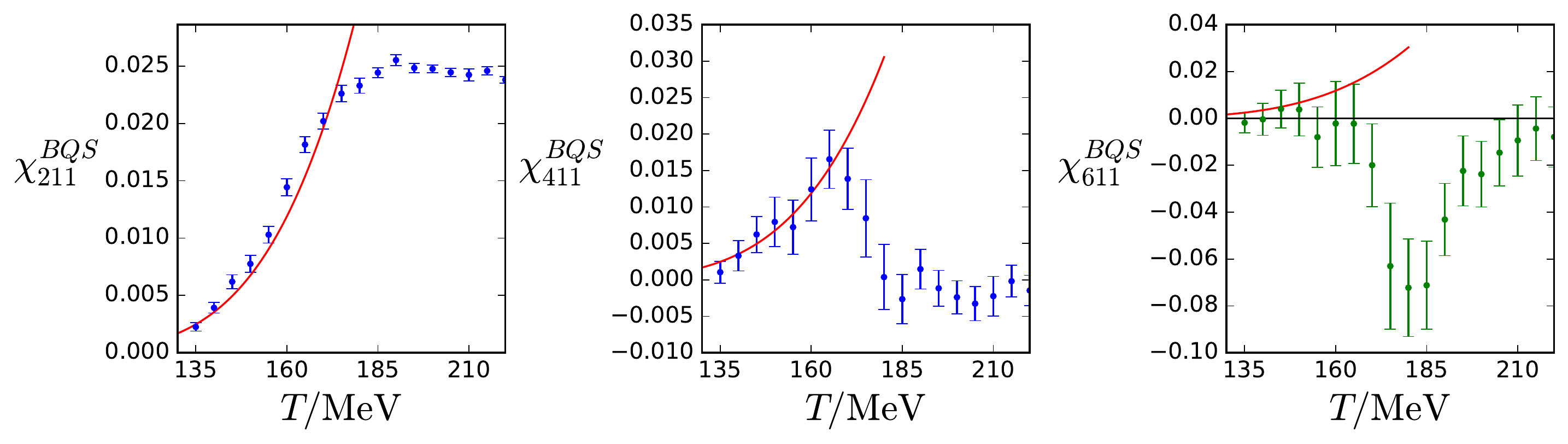}
\includegraphics[width = 0.9\textwidth]{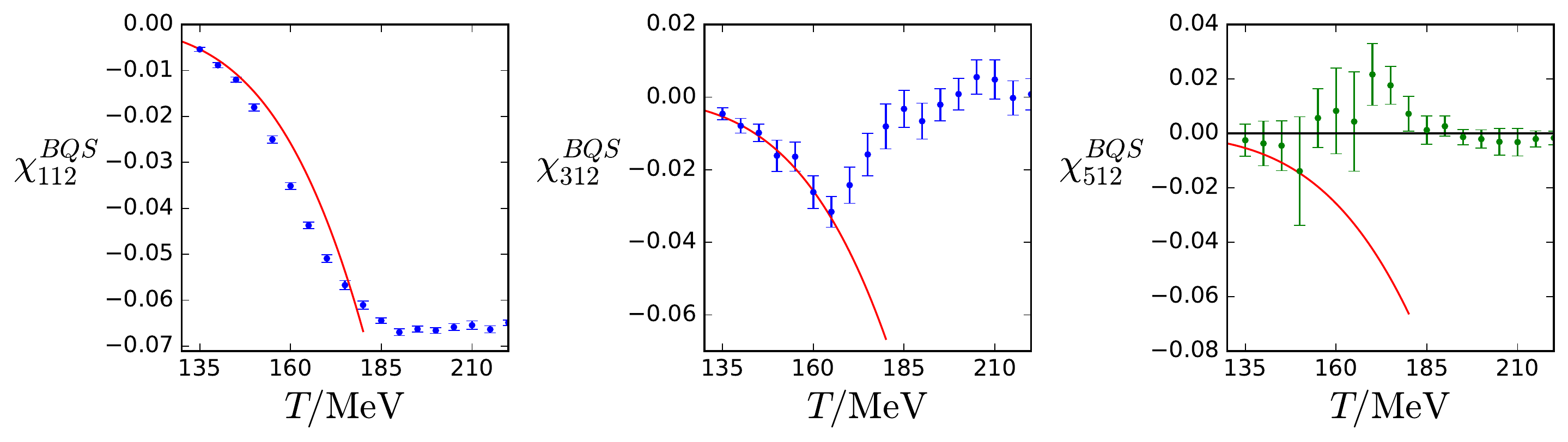}
\includegraphics[width = 0.9\textwidth]{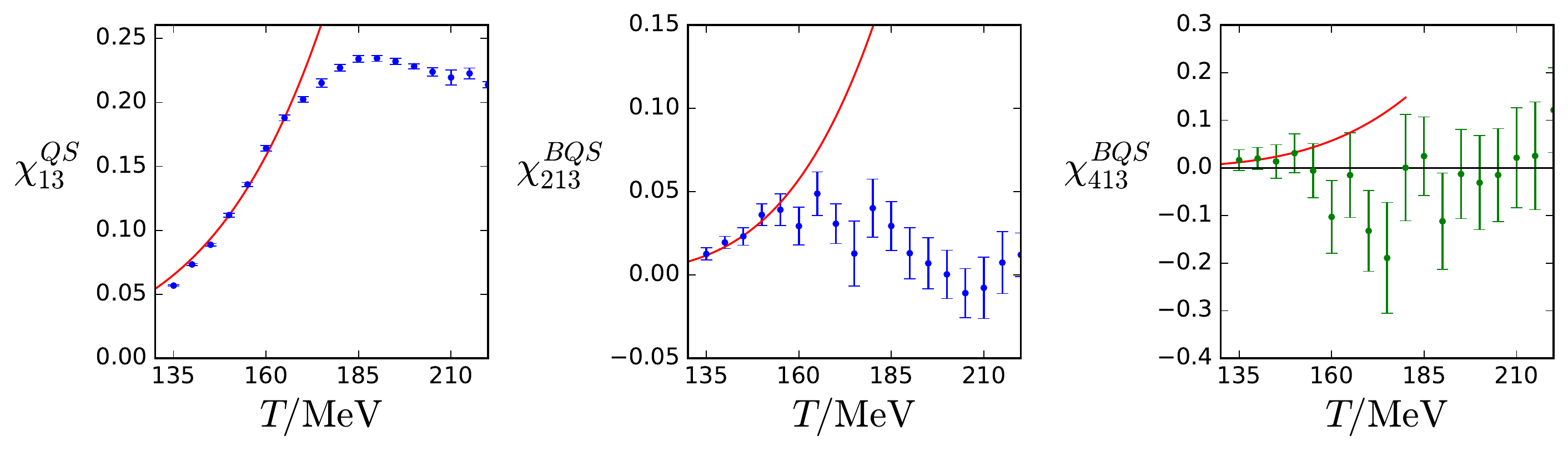}
 \end{center}
 \caption{Results containing one electric charge derivative
 on the various correlators on our $48^3\times12$ lattice as
functions of the temperature. Green data points denote our estimates for
the high orders, these were fitted using a prior distribution.
The red curves are the HRG model results.\label{resultQ1}}
\end{figure}

\begin{figure}[h!]
 \begin{center}
\includegraphics[width = 0.9\textwidth]{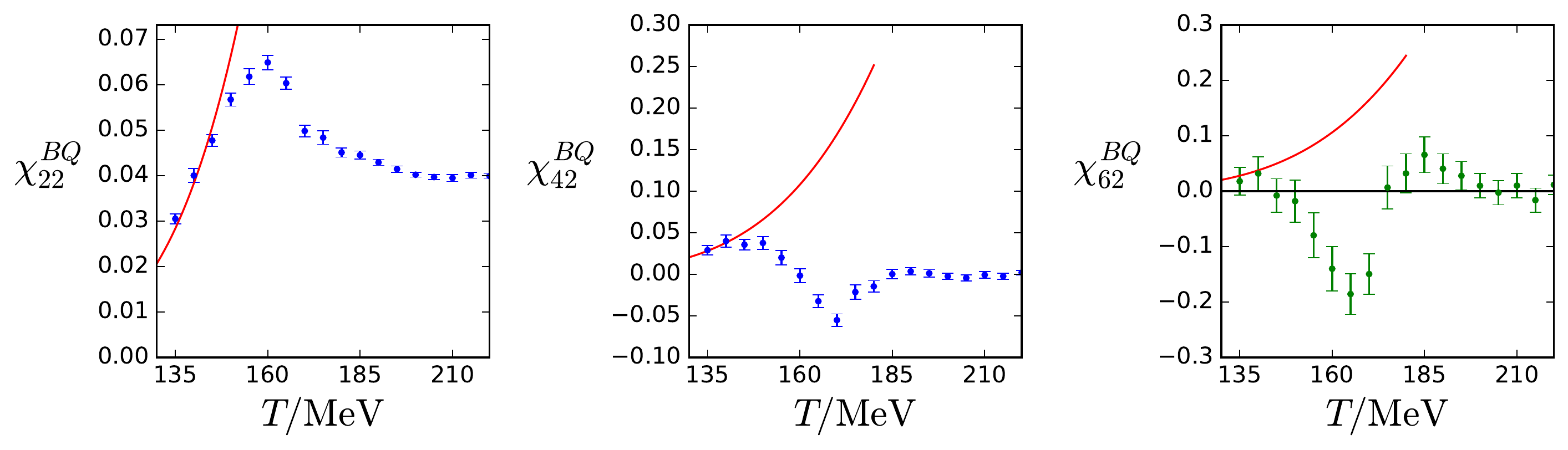}
\includegraphics[width = 0.9\textwidth]{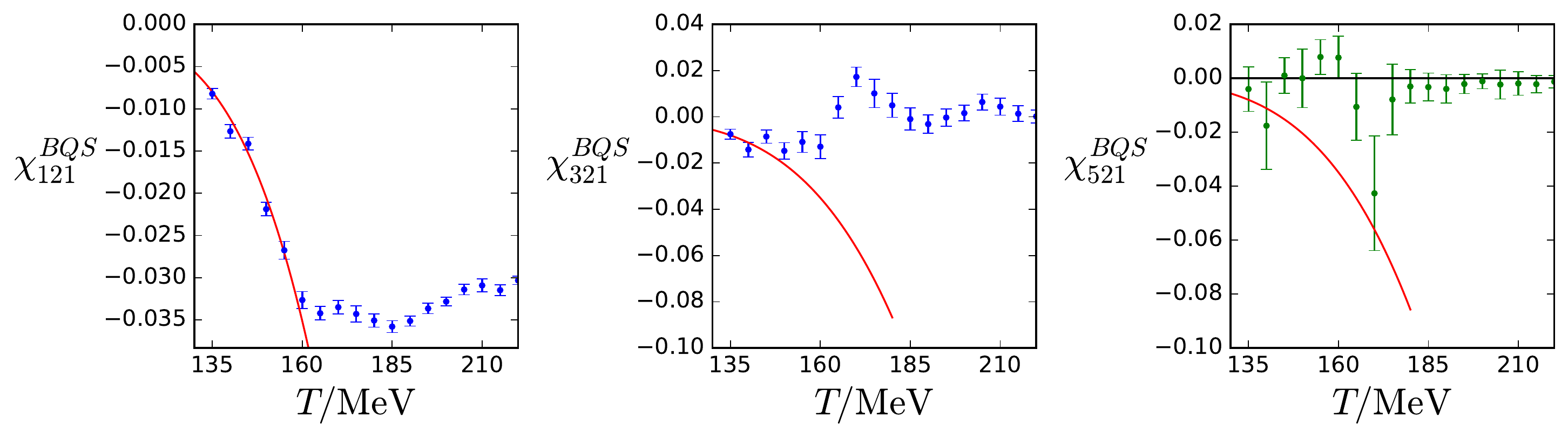}
\includegraphics[width = 0.9\textwidth]{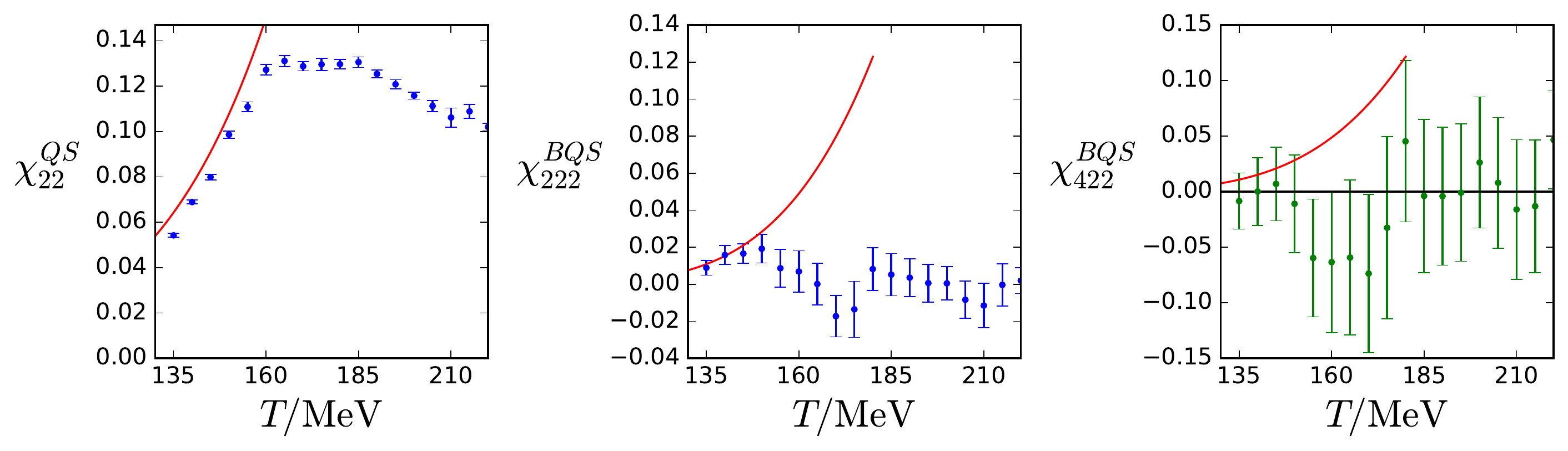}
 \end{center}
 \caption{Results containing two electric charge derivatives
 on the various correlators on our $48^3\times12$ lattice as
functions of the temperature. Green data points denote our estimates for
the high orders, these were fitted using a prior distribution.
The red curves are the HRG model results.\label{resultQ2}
Charge correlators without baryon derivative (here $\chi^{QS}_{22}$)
are expected to have significant discretization errors.}
\end{figure}

\begin{figure}[h!]
 \begin{center}
\includegraphics[width = 0.9\textwidth]{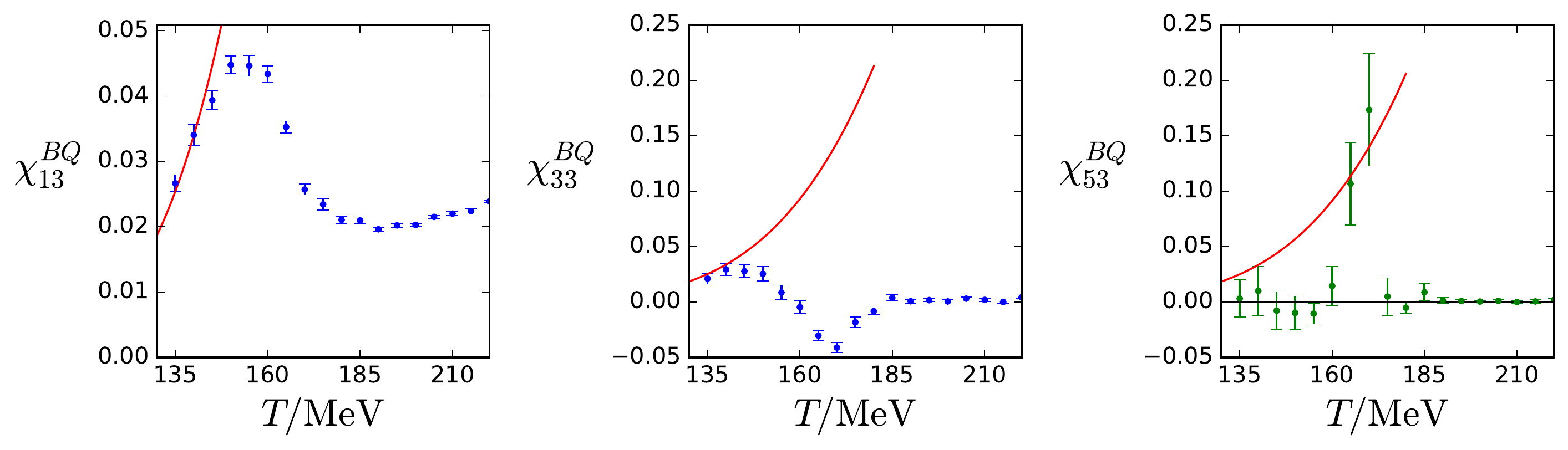}
\includegraphics[width = 0.9\textwidth]{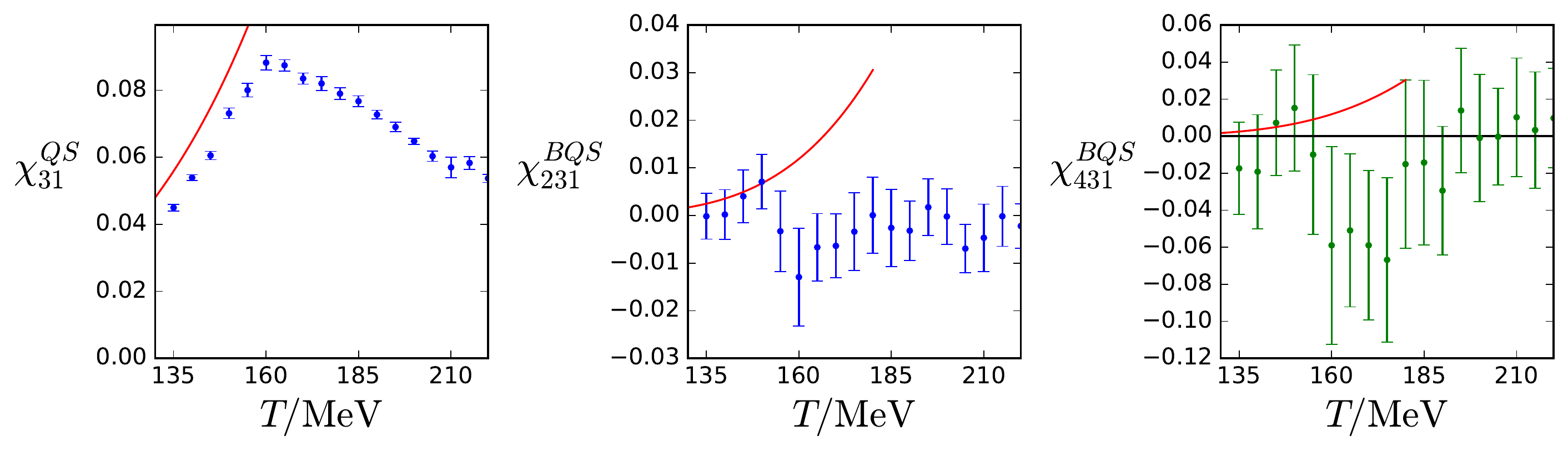}
 \end{center}
 \caption{Results containing three electric charge derivatives
 on the various correlators on our $48^3\times12$ lattice as
functions of the temperature. Green data points denote our estimates for
the high orders, these were fitted using a prior distribution.
The red curves are the HRG model results.\label{resultQ3}
Charge correlators without baryon derivative (here $\chi^{QS}_{31}$)
are expected to have significant discretization errors.
}
\end{figure}

\begin{figure}[h!]
 \begin{center}
  \includegraphics[width = 0.9\textwidth]{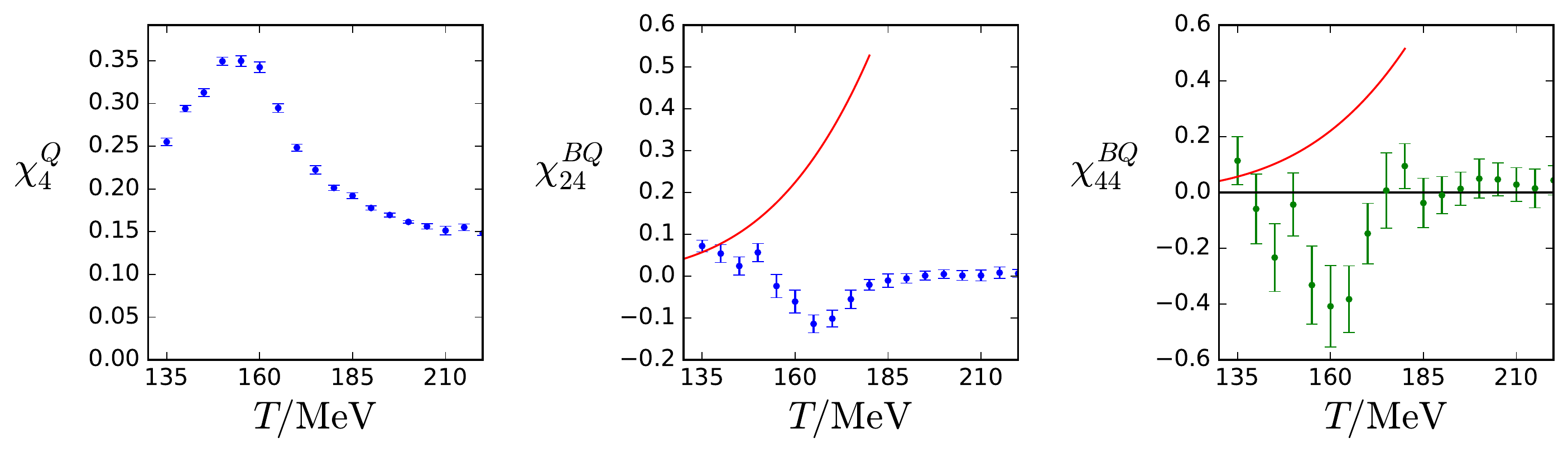}
 \end{center}
 \caption{$\chi_{4}^{Q},~\chi_{24}^{BQ}$ and estimate for $\chi_{44}^{BQ}$ as
functions of the temperature. The quantity $\chi_{4}^Q$ has severe cut-off effects
on this lattice \cite{Bellwied:2015lba}. The red curves are the HRG model results.
\label{resultQ4}}
\end{figure}

In the following plots we show these results organized by the number of
charge derivatives ($j$) in Figs.~\ref{resultQ0}-\ref{resultQ4}. It is notoriously
difficult to calculate charge correlators using staggered fermions \cite{Bellwied:2015lba}.
Correlators that are not protected by a baryon derivative are affected by
significant discretization errors.  It is understood in the HRG model context
that discretization errors mostly affect the contributions from pions and kaons.
Staggered lattice effects introduce the highest relative errors for the lightest mesons. 
Luckily, however, quantities with such discretization effects come with a small pre-factor into
the final formulas of Eqs.\.~(\ref{eq:1})-(\ref{eq:3}). If we had a complete
isospin symmetry (factor 0.5 between $\langle n_Q\rangle$ and $\langle n_B\rangle$ in
Eq.~(\ref{eq:neutrality})) then electric charge correlators would play no role at all
in the extrapolation of baryon fluctuations.

\section{Statistics and lattice details}

In Table~\ref{tab:stat} we give the number of analyzed configurations per ensemble.
The simulation parameters and the details of the analysis are given in Ref.~\cite{Bellwied:2015lba}.

The determination of the $\mu$ derivatives follows the lines of Ref.~\cite{Allton:2002zi,Bellwied:2015lba}.
We calculate four quantities per configuration and per quark mass
\begin{eqnarray}
A_j&=\frac{d}{d\mu_j} \log (\det M_j)^{1/4} = &\trt M_j^{-1} M_j'\,,\\
B_j&=\frac{d^2}{(d\mu_j)^2} \log(\det M_j)^{1/4} =&\trt\left(
M_j'' M_j^{-1}
-M_j' M_j^{-1} M_j' M_j^{-1}
\right)\,,\\
C_j&=\frac{d^3}{(d\mu_j)^3} \log(\det M_j)^{1/4} =&\trt\left(
M_j' M_j^{-1}
-3 M_j'' M_j^{-1} M_j' M_j^{-1}\right.\nonumber\\
&&
\left.
+2 M_j' M_j^{-1} M_j' M_j^{-1} M_j' M_j^{-1}
\right)\,,\\
D_j&=\frac{d^4}{(d\mu_j)^4} \log(\det M_j)^{1/4} = &\trt\left(
M_j'' M_j^{-1}
-4 M_j' M_j^{-1} M_j' M_j^{-1}
-3 M_j'' M_j^{-1} M_j'' M_j^{-1} \right.\nonumber\\
&&
\left.
+12 M_j'' M_j^{-1} M_j' M_j^{-1} M_j' M_j^{-1}\right.\nonumber\\
&&
\left.
-6 M_j' M_j^{-1} M_j' M_j^{-1} M_j' M_j^{-1}M_j' M_j^{-1}
\right)\,,
\end{eqnarray}
Here $M_j$ is the fermion matrix corresponding to the $j$-th quark mass in the system. $M'$ and $M''$ indicate
the first and higher order derivatives with respect to the quark chemical potential. For this simple staggered
action higher order derivatives are equal to lower order ones, $M'''=M'$ and $M''''=M''$ by construction.
These traces are calculated using the standard stochastic method, by calculating the effect of the matrices on random
sources.  At finite (imaginary) chemical potentials we used $4\times256$ Gaussian random sources for the light quarks
and $4\times128$ sources for the strange quarks. The analysis was accelerated by calculating 256 eigenvectors
of the Dirac operator first. These eigenvectors were then fed into an Eig-CG algorithm.

Using the isospin symmetry ($m_u=m_d$), the $ABCD$ traces can be used to calculate the $\chi^{uds}$ derivatives
with the following formulas:
\begin{eqnarray}
\chi^{uds}_{200} &=&
+ \langle B_u\rangle
+ \langle A_u^2\rangle
- \langle A_u\rangle^2
\\
\chi^{uds}_{110} &=&
+ \langle A_u^2\rangle
- \langle A_u\rangle^2
\\
\chi^{uds}_{101} &=&
+ \langle A_u A_s\rangle
- \langle A_s\rangle\langle A_u\rangle
\end{eqnarray}
\begin{eqnarray}
\chi^{uds}_{300} &=&
+ \langle C_u\rangle
+3 \langle A_u B_u\rangle
+ \langle A_u^3\rangle
-3 \langle B_u\rangle\langle A_u\rangle
-3 \langle A_u\rangle\langle A_u^2\rangle
+2 \langle A_u\rangle^3
\\
\chi^{uds}_{210} &=&
+ \langle A_u B_u\rangle
+ \langle A_u^3\rangle
- \langle B_u\rangle\langle A_u\rangle
-3 \langle A_u\rangle\langle A_u^3\rangle
+2 \langle A_u\rangle^3
\\
\chi^{uds}_{120} &=&
+ \langle A_u B_u\rangle
+ \langle A_u^3\rangle
- \langle B_u\rangle\langle A_u\rangle
-3 \langle A_u\rangle\langle A_u^2\rangle
+2 \langle A_u\rangle^3
\\
\chi^{uds}_{111} &=&
+ \langle A_u A_u A_s\rangle
- \langle A_s\rangle\langle A_u^2\rangle
-2 \langle A_u\rangle\langle A_u A_s\rangle
+2 \langle A_s\rangle\langle A_u\rangle^2
\end{eqnarray}
\begin{eqnarray}
\chi^{uds}_{400} &=&
+ \langle D_u\rangle
+3 \langle B_u B_u\rangle
+4 \langle A_u C_u\rangle
+6 \langle A_u^2B_u\rangle
+ \langle A_u^4\rangle\nonumber\\
&&
-4 \langle C_u\rangle\langle A_u\rangle
-3 \langle B_u\rangle^2
-6 \langle B_u\rangle\langle A_u^2\rangle
-12 \langle A_u\rangle\langle A_u B_u\rangle\nonumber\\
&&
-4 \langle A_u\rangle\langle A_u^3\rangle
-3 \langle A_u A_u\rangle\langle A_u^2\rangle
+12 \langle B_u\rangle\langle A_u\rangle^2\nonumber\\
&&
+12 \langle A_u\rangle^2\rangle\langle A_u^2\rangle
-6 \langle A_u\rangle^4
\\
\chi^{uds}_{310} &=&
+ \langle A_u C_u\rangle
+3 \langle A_u^2 B_u\rangle
+ \langle A_u^4\rangle
- \langle C_u\rangle\langle A_u\rangle
-3 \langle B_u\rangle\langle A_u^2\rangle\nonumber\\
&&
-6 \langle A_u\rangle\langle A_u B_u\rangle
-4 \langle A_u\rangle\langle A_u^3\rangle
-3 \langle A_u^2\rangle\langle A_u^2\rangle\nonumber\\
&&
+6 \langle B_u\rangle\langle A_u\rangle^2
+12 \langle A_u\rangle\langle A_u\rangle\langle A_u^2\rangle
-6 \langle A_u\rangle^4
\\
\chi^{uds}_{220} &=&
+ \langle B_u^2\rangle
+2 \langle A_u^2 B_u\rangle
+ \langle A_u^4\rangle
- \langle B_u\rangle^2
-2 \langle B_u\rangle\langle A_u^2\rangle\nonumber\\
&&
-4 \langle A_u\rangle\langle A_u B_u\rangle
-4 \langle A_u\rangle\langle A_u^3\rangle
-3 \langle A_u^2\rangle\langle A_u^2\rangle\nonumber\\
&&
+4 \langle B_u\rangle\langle A_u\rangle\langle A_u\rangle
+12 \langle A_u\rangle\langle A_u\rangle\langle A_u^2\rangle
-6 \langle A_u\rangle^4
\\
\chi^{uds}_{211} &=&
+ \langle A_u B_u A_s\rangle
+ \langle A_u^3A_s\rangle
- \langle A_s\rangle\langle A_u B_u\rangle
- \langle A_s\rangle\langle A_u^3\rangle
- \langle B_u\rangle\langle A_u A_s\rangle
- \langle B_u A_s\rangle\langle A_u\rangle\nonumber\\
&&
-3 \langle A_u\rangle\langle A_u^2 A_s\rangle
-3 \langle A_u A_s\rangle\langle A_u^2\rangle
+2 \langle A_s\rangle\langle B_u\rangle\langle A_u\rangle
+6 \langle A_s\rangle\langle A_u\rangle\langle A_u^2\rangle\nonumber\\
&&
+6 \langle A_u\rangle^2\langle A_u A_s\rangle
-6 \langle A_s\rangle\langle A_u\rangle^3
\end{eqnarray}
If the listed products of the $A,B,C,D$ traces are calculated as products of the stochastic
estimators, a bias could be introduced. Thus, in products different random
vectors have to be used in each factor. Alternatively, the expectation value of
the bias has to be subtracted.
The last step is to express the derivatives in terms of $\mu_B$, $\mu_Q$ and
$\mu_S$ in Eq.~(\ref{eq:chideriv}) using Eqs.~(\ref{eq:muBQS}), which is a
straightforward exercise.

\begin{table}[ht]
\begin{tabular}{|c|c|c|c|c|c|c|c|c|}
\hline
$T$ [MeV] &${\hat\mu_B}^I=0$&${\hat\mu_B}^I=0.4$&${\hat\mu_B}^I=0.8$&${\hat\mu_B}^I=1.2$&${\hat\mu_B}^I=1.6$&${\hat\mu_B}^I=2.0$&${\hat\mu_B}^I=2.4$&${\hat\mu_B}^I=2.7$\\\hline
135 &17871&1647&2680&4377&2375&3449&2622&2008\\
140 &22624&1625&3583&2975&3499&5321&3129&3211\\
145 &17195&2439&5255&4468&3191&2846&4959&4117\\
150 &18429&2048&3404&10115&6450&5665&3211&3254\\
155 &17494&1624&4735&4938&3911&7813&3670&3485\\
160 &12688&1607&4459&4831&3382&3917&4831&4990\\
165 &18472&1935&4976&8113&8466&4984&5235&4321\\
170 &14417&1987&2704&8820&8053&8023&5916&3273\\
175 &12018&2034&2006&4748&3878&11330&6178&5583\\
180 &12446&2104&2089&5424&4514&6057&5910&4466\\
185 &14184&2151&2138&3112&3086&5934&7733&3767\\
190 &13741&1693&3395&4395&8140&10410&4201&3844\\
195 &15013&1758&3643&5334&8420&5707&3884&4003\\
200 &14974&2300&2262&5999&10709&5033&5496&4203\\
205 &7788&2126&2125&5951&5873&8294&3087&4333\\
210 &4014&1957&1949&12174&6649&3543&2999&3146\\
215 &2506&1783&7056&2268&2244&1711&1674&2090\\
220 &9172&1810&3548&4264&5498&1754&1717&2163\\
\hline\end{tabular}
\caption{\label{tab:stat}Statistics of our simulations on the $48^3\times12$ lattice.
We list the number of stored and analyzed gauge configurations. These configurations
were separated by ten Rational Hybrid Monte Carlo updates.
}
\end{table}

\clearpage
\bibliography{thermo}
\bibliographystyle{JHEP}
% --------------  Ende Text----------------------

\clearpage
\vfill

\end{document}